# From IBD tracts to runs of homozygosity: a unified coalescent framework including selection

Enrique Santiago

Departamento de Biología Funcional, Facultad de Biología, Universidad de Oviedo, 33006 Oviedo, Spain.

ORCID: 0000-0002-5524-5641

# Abstract

Identity-by-descent (IBD) tracts and runs of homozygosity (ROH) represent the underlying and observable aspects of chromosomal autozygosity, but no formal relationship between them had been established. This study develops a coalescent framework that connects these concepts, providing tools for detecting selection signatures and extending the temporal range of effective population size ($N_e$) inference. Closed-form probability functions are derived for IBD tract lengths and mapped onto observable ROHs by modelling the displacement of autozygosity boundaries from true recombination breakpoints to the nearest heterozygous marker. This mapping explicitly accounts for mutation, gene conversion, marker density, and heterozygosity. Selection effects are incorporated analytically, yielding computational tools to predict tract length distributions under background selection and partial or complete selective sweeps. Background selection shifts IBD tract-length distributions toward longer tracts, an effect better explained by a generation-dependent $N_e$ reduction below its neutral value than by a single constant $N_e$. Selective sweeps generate directional asymmetry in tract-length distributions at neutral sites linked with the selected locus, with its direction indicating on which side of the neutral site the selected locus resides. This asymmetry decays rapidly after the sweep, whereas the local $N_e$ reduction estimated from the median tract length persists considerably longer. The coalescent framework is further extended to arbitrary time-varying population size, enabling insights into past demography across a wide temporal range.

## Article Summary

Genome sequences show runs of homozygosity, while coalescent theory describes identity-by-descent tracts. Here the two concepts are connected explicitly, with mutation, gene conversion, and marker density appearing as parameters of the tract-length distribution rather than as corrections applied afterwards. This connection renders short tracts more usable and extends demographic inference further into the past. The same framework predicts how background selection and selective sweeps distort tract lengths, and why selection is better described by a time-varying effective size than by a single reduced value.

# Introduction

Identity-by-descent (IBD) tracts and runs of homozygosity (ROH) are genomic manifestations of autozygosity. An IBD tract is a contiguous chromosomal segment shared by two haplotypes descending from a single ancestral copy with no internal recombination (Thompson 2018), whereas an ROH is a homozygous stretch identified by typed markers within a diploid individual (McQuillan *et al.* 2008). True IBD tract boundaries are set by unobserved recombination breakpoints, whereas ROH boundaries are localized by flanking heterozygous markers. Because autozygosity is shaped by population demography, mating patterns and selection (Ceballos *et al.* 2018), IBD tracts and ROH inform about effective population size ($N_e$) and genomic signatures of selection. Recombination shortens autozygosity tracts over generations, creating the inverse relationship between tract length and coalescence time that links their distributions to demographic and selective history.

IBD tract-length theory has evolved from classical genetics to coalescent-based demographic inference. Fisher (1949, 1954) developed the theory of junctions to derive the expected autozygosity under regular mating systems. Stam (1980) and Chapman and Thompson (2003) extended this theory to finite random-mating populations. Sved (1971) introduced the concept of chromosome segment homozygosity (CSH), later applied to $N_e$ estimation by Hayes *et al.* (2003). Advances in sequencing and IBD detection (Browning and Browning 2013; Seidman *et al.* 2020; Guo *et al.* 2025) subsequently enabled its widespread use in demographic inference (Palamara *et al.* 2012; Browning and Browning 2015; Fournier *et al.* 2023). However, these developments did not yield a unified probability model connecting theoretical IBD tracts to the observed ROH distributions under finite marker density.

Inferring IBD tracts from genomic data is challenging because recombination boundaries must be reconstructed from identity-by-state patterns at a resolution determined by marker density and heterozygosity. The combined effect of *de novo* mutations and gene conversion introduce artefactual breakpoints within otherwise identical haplotypes (Browning and Browning 2024), whereas conflation can merge adjacent IBD segments from different ancestors into a single longer segment (Chiang *et al.* 2016). Both corrupt the short end of the tract-length spectrum, that is the range that carries the information about the distant past. This is particularly relevant in the context of the low marker density characteristic of samples from natural populations (Shafer and Kardos 2025).

ROH-based approaches address these difficulties by analyzing the observed runs directly. MacLeod *et al.* (2009, 2013) predict the probability that two homologous chromosomes remain identical by state over a given distance and infer piecewise demographic histories by matching predicted and observed ROH distributions.

Pemberton *et al.* (2012), by contrast, classify ROH into length groups reflecting broad demographic processes. Because in both cases the history is obtained without an explicit probability density for the ROH length spectrum, there is no way to separate what the data determine and what the fitting procedure imposes: how many epochs are resolvable, where their boundaries can be placed, and how far into the past the length distribution still carries information.

Selection leaves its own imprint on tract lengths: a selective sweep carries a haplotype to high frequency faster than recombination can erode it, so IBD tracts near the selected site are longer and more frequent than the average in the genome (Albrechtsen *et al.* 2010; Browning and Browning 2020). Scans for selection exploit this, marking regions where tract length, coverage or frequency exceeds the genome background or neutral simulations (Sabeti *et al.* 2002; Guo *et al.* 2024), and Temple *et al.* (2024) went further, fitting inferred tract-length distributions around known adaptive variants to estimate their selection coefficients in human populations. What is still missing is the distribution expected at a neutral site at a given distance from the selected one: without it there is no analytical baseline against which an excess can be judged, and no way to tell from the lengths alone where the target of selection lies.

Building on the relationship between tract length and coalescence time, this study develops a theoretical framework linking IBD tract-length distributions to observable ROH distributions. Closed-form IBD tract-length densities are derived and mapped to ROH distributions while accounting for finite marker density, variable heterozygosity, de novo mutation, and gene conversion. Under arbitrary time-varying population size, tract-length distributions are obtained by integrating over the coalescence-time distribution, with closed-form solutions for piecewise-constant demographic histories. The framework also incorporates background selection (Charlesworth *et al.* 1993; Santiago and Caballero 2016) and partial or complete selective sweeps (Gillespie 2000) into coalescent predictions at neutral sites.

# Materials and methods

## - Model assumptions

A Wright-Fisher model of $2N$ haploid individuals reproducing at random is assumed throughout. $N$ denotes the effective population size in the absence of selection. $N_e$ is reserved for the effective size reduced below $N$ when selection is considered, in the sense of the $N_e/N$ ratio of classic background-selection theory (Charlesworth *et al.* 1993). The genome is a continuous sequence of sites, in which a neutral focal site is inserted at a random position. This focal site is virtual and has no associated genotype. For two haploid genomes that are identical by descent (IBD) at the focal site, the probability that they both trace back to a single ancestral copy $t$ generations in the past is given by:

$$f(t) = \frac{1}{2N}\left(1 - \frac{1}{2N}\right)^{t-1} \approx \frac{e^{-t/2N}}{2N}$$

At each meiosis, three kinds of events interrupt the continuity of the original chromosome on either side of the focal site: recombination occurring at a rate of $c$ and the combined effect of *de novo* mutations and gene conversions at a rate $m$. These rates are given per Mb, thus, for a genome with a constant recombination rate of 1 cM/Mb there are $c = 0.01$ recombination events per Mb.

## - Simulations and analysis

A coalescent simulation of time-varying demography was performed with msprime v1.4.4 (Baumdicker *et al.* 2022), using the demographic trajectory shown in Figure 7. The genome comprised ten 135-Mb chromosomes with uniform per-site recombination and neutral mutation rates of $10^{-8}$ and $2.5\ 10^{-8}$, respectively, yielding a mean heterozygous-marker spacing of $D/H = 5.2\ 10^{-4}$ Mb. The trajectory shown in Figure 7 was inferred with the program Ne_trajectory.py (see Data and Software Availability), which implements the theory developed in Appendix Section 5.

Simulations involving selection or varying marker density used custom forward-time programs. In all simulations, ROH length distributions were calculated from distances between consecutive heterozygous sites in samples of 200 individuals and grouped into 15-kb bins. Unless otherwise specified, a burn-in of $10^4$ generations was used to approach mutation-selection-drift equilibrium.

Marker-density effects were simulated in genomes comprising a single 200 Mb chromosome. Three diploid population sizes were considered: 100, 1,000, and 10,000 individuals. The recombination rate was $c = 0.01$ events per Mb, $D/H$ ranged from 0.0128 to 0.2048 Mb, and mutation rates from 0.00012 to 0.02026 per Mb, scaled to obtain the desired equilibrium marker density at each population size. For each parameter combination, genomic ROH coverage was calculated by length bin, and $N$ was estimated for each bin using Equation 9, with the finite-marker-density correction, and Equation 4, without it. Estimates based on the median of the tract length distribution ($p_{site}(x) = 0.5$) were also calculated using Equation A26, with the correction, and Equation 6, without it. All estimates were averaged across four replicates to generate Figure 2.

Background selection was simulated in a population of 1,000 diploid individuals carrying a single 200 Mb chromosome, with uniform per-site recombination and neutral mutation rates of $10^{-8}$ and $3.3\ 10^{-7}$ respectively. Deleterious mutations occurred at a rate of 0.01 per Mb, with a multiplicative deleterious effect of 0.02 per copy. After the burn-in period, the observed ROH length distribution across the entire genome was calculated. The program PDF_IBD_tracts_BS.py (see Data and Software Availability) used Equation 10 to predict the ROH length distribution and generate Figure 3.

Selective sweeps were simulated in 4-Mb regions with equal recombination, gene conversion and neutral mutation rates of $10^{-8}$ events per-site, respectively, in populations of 15,000 diploid individuals. After a burn-in of $3\ 10^{4}$ generations, a single copy of a dominant beneficial allele with a selection coefficient of 0.05 was introduced at the center of the region. Replicates in which the allele was lost were discarded. For Figure 4, the simulation continued until the beneficial allele reached a frequency of 90% in one replicate. The distribution of ROH lengths was then calculated on either side of a neutral focal site located at a distance of 0.2 Mb from the selected site. The predicted distribution was calculated solving Equation 11 with the program PDF_IBD_tracts_SS.py (see Data and Software Availability). For Figure 5, samples were taken at different stages of the sweep, and the ROH length distributions were calculated at focal sites spaced 3.5 kb apart across the region. At each focal site, two $N_e$ estimates were obtained: one from Equation A26 evaluated at $p_{site}(x) = 0.5$ and the other from Equation 5. The left-right asymmetry was calculated as the difference in cM between the means of the ROH length distributions on left and right sides of each position. To minimize edge effects caused by the limited size of the simulated region, the figures show only the central 3 Mb.

## - Human genomic data and analysis

$N_e$ and the left-right asymmetries were estimated across two 3-Mb regions of the human genome encompassing two genes associated with local adaptations: lactase persistence (*MCM6*) and skin pigmentation (*SLC24A5*). Six populations representing the presence and absence of these adaptations were selected from the 1000 Genomes Project database (expanded Phase 3 high-coverage panel; release 20220422; GRCh38;

ENA accession PRJEB55077): YRI (108 individuals) and ESN (99) from Africa, GBR (91) and CEU (99) from Europe, and CHB (103) and CHS (105) from East Asia. ROH length distributions were then calculated using the unfiltered genome sequence, with focal sites distributed at 3.5 kb intervals. Sex-averaged maps were used for the three processes that break the identity: recombination (Halldorsson *et al.* 2019), gene conversion (Palsson *et al.* 2025) and mutation (Chen *et al.* 2024). Local estimates of $N_e$ corresponding to Figure 6 were obtained at intervals of 3.5 kb by two methods: using Equation A26 evaluated at the median of the tract length distribution ($p_{site}(x) = 0.5$), and Equation 5 based on the mean tract length. The asymmetry was calculated as the difference between the means in cM of the observed ROH length distributions at the left and right sides of each position.

# Results

The theory is developed in six steps. First, the two tract length distributions are re-derived on a physical map for constant-size populations: the distribution conditional on coalescence time and the reverse conditional distribution, i.e. coalescence time on tract length. The estimators of effective population size then follow from these distributions. The remaining four steps relax one assumption at a time: finite marker density, background selection, selective sweeps, and a population size that varies with time.

## - The two IBD tract-length distributions conditional on coalescence time

Two genomes that are identical by descent at a focal site will share flanking segments of the same ancestral chromosome. Break events accumulated along each genealogical branch independently truncate these copies on either side of the focal site. The overlapping region defines the IBD tract associated with that site (Figure 1). Here, IBD is used in a broad sense because tract continuity may be disrupted not only by recombination at rate $c$, but also by mutation and gene conversion at a combined rate $m$.

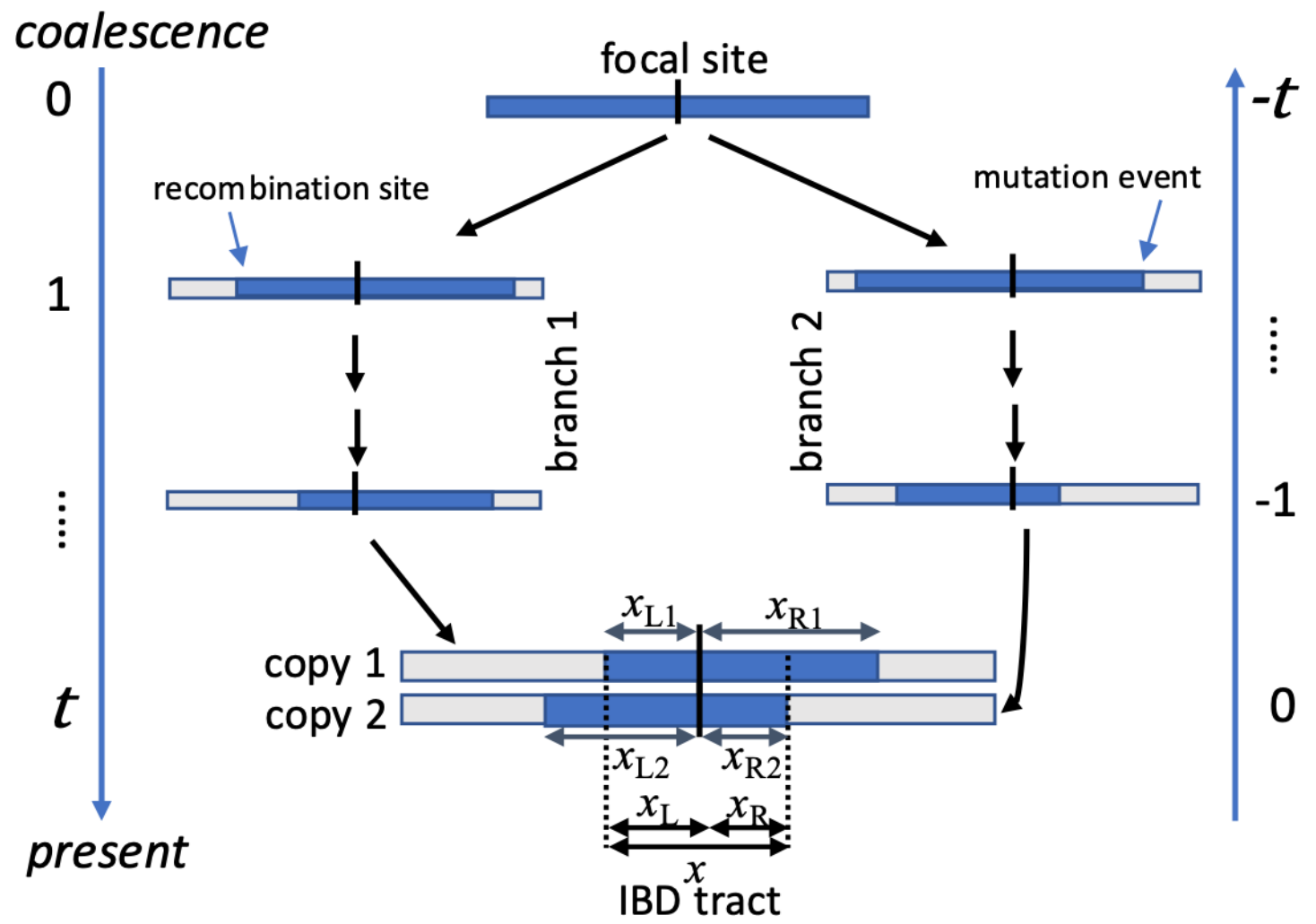

**Figure 1**. An IBD tract of length $x$ Mb at the present generation, $t$ generations after coalescence. Recombination, *de novo* mutation and gene conversion events trim both copies of the original chromosome on either side of the focal site. In equations, the subscripts of $N$ with negative time indexes represent population size at generations back in the past.

Because recombination, mutation, and gene-conversion events occur as independent Poisson processes, the distance between consecutive breakpoints within a generation follows an exponential distribution with per-Mb rate $B = c + m$. As these events accumulate independently over generations, the conserved tract lengths on either side of the focal site (i.e. $x_{R1}$, $x_{R2}$, $x_{L1}$ and $x_{L2}$ measured in Mb) at generation $t$ are exponentially distributed with rate $tB$. The right boundary of the IBD tract, $x_R$, is the shorter of the two branch-specific distances, $x_{R1}$ and $x_{R2}$. This reflects that breakpoints accumulate along both branches, resulting in an exponential distribution with rate $2tB$:

$$P(x_R|t) = 2tB \cdot e^{-2tBx_R} \tag{1}$$

Since the left and right flanking lengths are independent and identically distributed, the total tract length, $x = x_R + x_L$, follows an Erlang distribution with shape 2 and the same rate as the two exponential distributions,

$$P(x|t) = (2tB)^2 \cdot x \cdot e^{-2tB \cdot x} \tag{2}$$

with mean $1/(tB)$ Mb.

Equation (2) applies to tracts that span the focal site. Because the probabilities of randomly selected tracts spanning the focal site is proportional to its length, these tracts constitute a length-biased subsample of all IBD tracts in the genome. Removing this bias by weighting Equation 2 by the inverse tract length gives an overall length distribution proportional to $P(x|t)/x$. After normalization, this distribution reduces to the exponential form in Equation (1),

$$Q(x|t) = 2tB \cdot e^{-2tBx}$$

with a mean tract length half that of tracts spanning the focal site: $1/(2tB)$ Mb.

## - Equations for coalescence time conditional on IBD tract length

Although the distribution of IBD tract lengths conditional on coalescence time is independent of $N$, the distribution of coalescence times depends on $N$. Combining the IBD tract-length distribution with the pairwise coalescence probabilities at the focal site across past generations yields a useful set of equations for estimating $N$ from the observed distribution of IBD tract lengths.

For a constant $N$, the coalescence time conditional on a tract of length $x$ follows an Erlang distribution with shape parameter 3 and rate parameter $\varphi = 2xB + 1/(2N)$ (see derivation of Equation A6 in the Appendix). Thus, the expected coalescence time for tracts of length $x$, regardless of whether they span the focal site, is given by:

$$E(t) = \frac{3}{2xB + \frac{1}{2N}} \text{ generations} \tag{3}$$

Any Erlang distribution with shape parameter 3 has a coefficient of variation of $1/\sqrt{3} = 0.577$, and its 90% central interval extends from $0.82/\varphi$ to $6.30/\varphi$ generations, irrespective of tract length. For long tracts, the expected coalescence time becomes independent of $N$ and approaches $3/(2xB)$ generations. For very short tracts, it approaches $6N$ generations.

Averaging the expectations in Equation 3 yields an expected coalescence time of $2N$ generations for a tract randomly sampled among those spanning the focal site:

$$\int_0^\infty P(x|t) \cdot E(t) \cdot dx = 2N \text{ generations}$$

By contrast, a tract randomly sampled from the entire genome, irrespective of whether it spans the focal site, has twice this expected coalescence time:

$$\int_0^\infty Q(x|t) \cdot E(t) \cdot dx = 4N \text{ generations}$$

At mutation-recombination-drift equilibrium in the absence of selection, the steady-state distribution $P(x)$ of IBD tract lengths spanning the focal site can be obtained by integrating Equation (2) over all past generations (see derivation of Equation A8):

$$P(x) = \frac{4xB^2}{N\left(2xB + \frac{1}{2N}\right)^3}$$

This equation also gives the density of genomic coverage by IBD tracts of length $x$. Dividing the focal-site density by $x$ yields the genome-wide tract length density, and multiplying the latter by $x$ recovers the genomic coverage. Therefore, the proportion of the genome covered by IBD tracts with lengths in the small interval $x \pm h/2$ is:

$$p(x;h) = h \cdot \left(P(x) \cdot \frac{1}{x}\right) \cdot x = h \cdot P(x) \tag{4}$$

Palamara *et al.* (2012) and Browning and Browning (2015) derived this equation, although in genetic-map units and without accounting for mutation or gene conversion. It directly yields two complementary estimators of $N$ at a focal site. The first uses the mean length, $\bar{x}_{site}$ of IBD tracts spanning the site (see the derivation of Equation A9 in the Appendix):

$$\bar{x}_{site} = \frac{\ln(2N)}{2NB} \text{ Mb} \tag{5}$$

The result for $B = 1$ closely resembles that derived by Sved (1971). The difference reflects the treatment of recombination: Sved did not account for the full distribution of recombination events along the chromosome, whereas the present derivation assumes an infinitely long chromosome.

The second estimator uses the proportion $p_{site}(x)$ of IBD tracts spanning both the focal site and a second site $x$ Mb away (see derivation of Equation A10 in the Appendix):

$$p_{site}(x) = \frac{1}{2N(e^{2xB} - 1) + 1} \approx \frac{1}{4NxB + 1} \tag{6}$$

Whereas Equation 5 integrates information from break events across the entire chromosome, Equation 6 considers only those occurring within a distance of $x$ Mb from the focal site, making it more sensitive for detecting local footprints of selection.

## - Connecting IBD and ROH concepts

Recombination, mutation, and gene-conversion events define the true boundaries of IBD segments. In practice, however, finite marker density causes empirically detected ROHs to deviate from these underlying segments. The model accounts for this discrepancy by explicitly incorporating uncertainty in boundary displacement (see Section 8 of the Appendix).

When a recombination event occurs in a branch, within the segment that still conserves the identity (blue chromosome segments in Figure 1), the observed ROH boundary shifts from the true recombination breakpoint to the nearest *het* site, defined as a site at which the two branches carry different alleles (see Section 8 in the Appendix). This displacement, which applies only to recombination events, follows an exponential distribution with rate $H/D$, where $H$ is the mean heterozygosity of segregating sites and $D$ is their mean physical distance in Mb. The distance to the observed boundary is the sum of the distance to the recombination breakpoint and its displacement to the nearest *het* site. Therefore, this sum follows a hypoexponential distribution with rates $c$ and $H/D$.

Conditional on coalescence times, the distribution of ROH lengths on one side of the focal site is approximated by:

$$P(x_R|t) \approx 2tB \cdot e^{-2txB}\left[\frac{1}{1-\gamma}\right]^{2t} \tag{7}$$

where $\gamma = cD/H$. Accordingly, the conditional distribution of the total length $x$ of an ROH spanning the focal site is:

$$P(x|t) \approx 4t^2B^2 \cdot x \cdot e^{-2txB}\left[\frac{1}{1-\gamma}\right]^{4t} \tag{8}$$

The fraction of the genome covered by ROHs with lengths within a small window of width $h$ centered $x$ Mb is approximates by:

$$p(x;h) \approx h \cdot \frac{4xB^2}{N\left(2xB + \frac{1}{2N} - 4\gamma\right)^3} \tag{9}$$

Equations 7-9 are not true probability density functions because they apply only to tract lengths $x \gg D/H$. As the marker density approaches infinity (i.e. $D \to 0$), the displacement term vanishes and all three expressions become equivalent to their IBD counterparts Equations 1, 2, and 4 respectively.

These predictions of the ROH length distribution rely on the simplifying assumption that heterozygous sites are randomly distributed along chromosomes. The consequences of this simplification were evaluated by measuring the bias caused on estimates of effective population size (Figure 2).

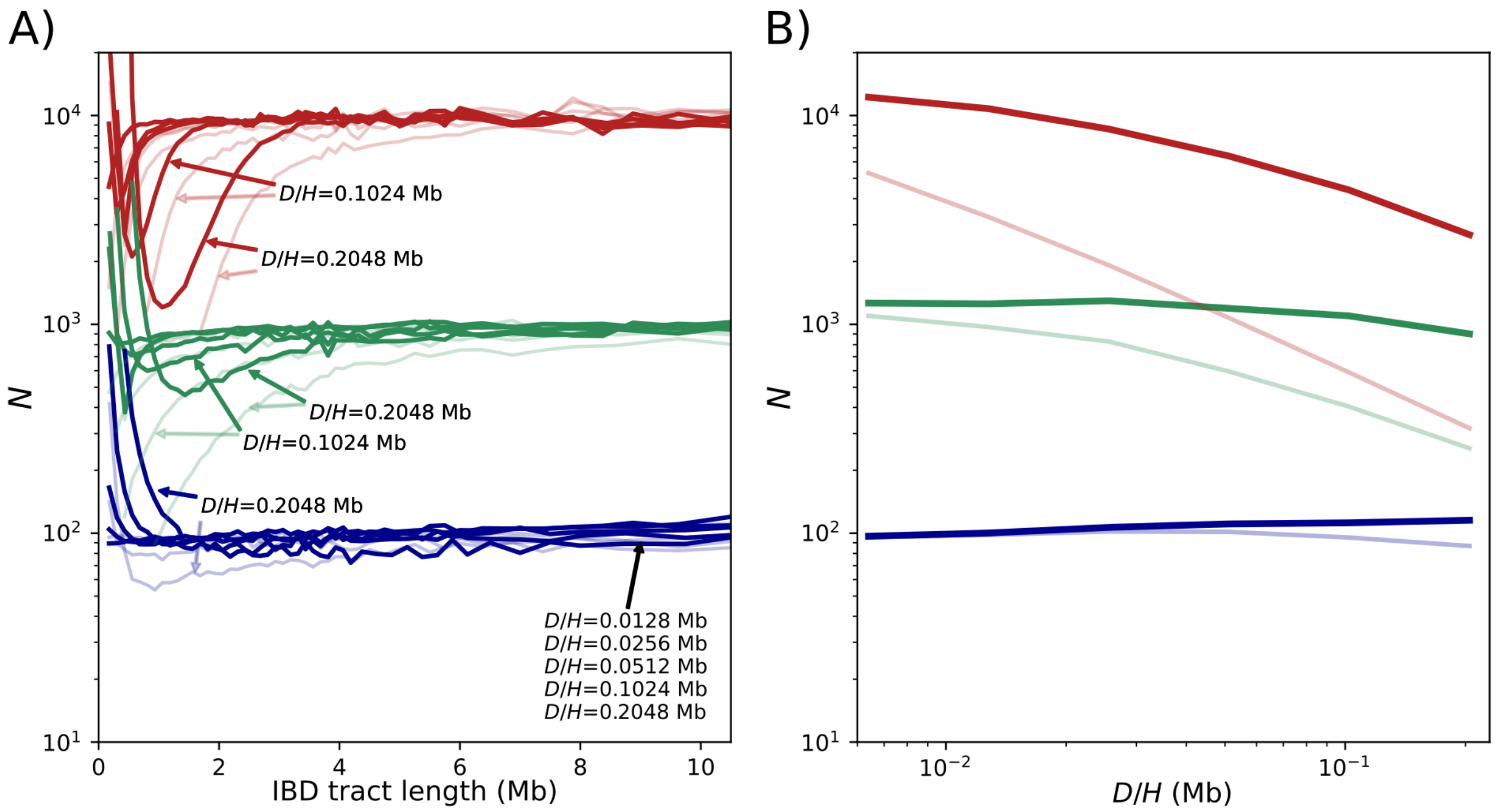

**Figure 2**. Improvements in accuracy of effective population size estimates using corrections for finite marker density. (A) Estimates of $N$ derived from the fraction of the genome covered by ROHs within each length bin, using Equation 9 with the finite-marker-density correction (dark lines) or Equation 4 without it (pale lines). (B) Estimates derived from the median length of ROHs spanning a focal site, using Equation A26 for the median ($p_{site}(x) = 0.5$) with the correction (dark lines) or Equation 6 without it (pale lines). Colors indicate population size: 100 (blue), 1,000 (green), and 10,000 (red) diploid individuals. Five marker-density conditions were considered, with average distance between heterozygous sites ($D/H$) ranging from 0.0128 to 0.2048 Mb. The recombination rate was $c = 0.01$ per Mb (i.e. 1 cM/Mb).

At high marker density ($D/H$<0.0512 Mb), the corrected estimators closely recovered the true $N$: Equation 9, based on ROH coverage, remained accurate across nearly the entire range of tract lengths, while Equation A26, based on the median ROH length (i.e. $p_{site}(x) = 0.5$), produced similar accurate estimates. The main exception involved tracts shorter than 1 Mb in populations of 10,000 individuals. Accuracy declined as marker density decreases (i.e. as $D/H$ increases), with the strongest bias occurring for short tracts and large populations. Thus, reliable inference of $N$ from short ROHs, which predominate in large populations, requires sufficiently density of markers.

## - The distribution of IBD tract lengths under background selection

Selection has a cumulative effect on genetic drift because associations between selected and neutral alleles persist across generations, even under free recombination (Robertson 1961; Santiago and Caballero 1995). Because fitness advantages are transmitted from parents to descendants, changes in the frequencies of associated neutral alleles become correlated across generations, increasing drift. This effect can be represented as a progressive reduction in $N_e$ over successive generations (Santiago and Caballero 1998): older neutral mutations experience stronger cumulative drift than younger ones (see Section 6 of the Appendix). Under background selection with multiplicative fitness effects, this time-dependent reduction in $N_e$ is given by the following equation (Santiago and Caballero 2016):

$$N_{e(t)} = N \cdot e^{-\frac{2}{cL}\int_0^{cL/2} V_w \cdot Q_{r(t)}^2 dr}$$

where $Q_{r(t)} = \sum_{i=0}^{t}(1-r)^i \cdot \left(1-\frac{V_M}{V_w}\right)^i$ represents the cumulative effect of selection over $t$ generations forward in time; $L$ is the chromosome length in Mb, so that, at a constant rate of $c$ recombination events per Mb, $cL$ is its length in morgans; $V_w$ is the standing genetic variance for fitness (i.e. the variance of the expected contributions of individuals to the next generation), and $V_M$ is the input of new genetic variance for fitness per generation. $N_{e(t)}$ is relative to a time point $t = 0$ in the past corresponding to the origin of a new neutral mutation and cannot be assigned to a particular calendar generation in a population.

The equation applies in reverse to coalescence processes and negative subscripts refer to generations in the past. The probability of coalescence $t$ generations ago is $f(t) = \frac{1}{2N_{e(-t)}} \cdot \prod_{i=1}^{t-1}\left[1-\frac{1}{2N_{e(-i)}}\right]$, where the subscript -1 in $N_e$ refers to the generation immediately preceding the present and the subscript -$t$ refers of the generation of coalescence. The $N_e$ values decrease backwards in time until the generation of coalescence is reached. Combining these coalescence probabilities with the PDF of IBD tract lengths for $P(x|t)$ (Equation 2) yields the proportion of the genome covered by IBD tracts of lengths within the small interval $x \pm h/2$ Mb (see derivation in Appendix, Section 6):

$$p(x;h) = h \cdot \sum_{t=1}^{\infty}[P(x|t) \cdot f(t)] \tag{10}$$

The fitting of this equation to observed values from simulation is good even under strong purifying selection (Figure 3). The distribution of IBD tracts accumulates at very short lengths in the absence of selection, while background selection shifts this distribution towards longer tracts. The reduction in effective population size across the genome caused by the cumulative effect of deleterious mutations at linked sites increases the rate of coalescence, thereby inflating the apparent tract lengths. As the impact of selection on drift accumulates over time, recent coalescence events are associated with a higher $N_e$ than ancient ones. This creates a distortion of the distribution that cannot be captured by a constant $N_e$ model.

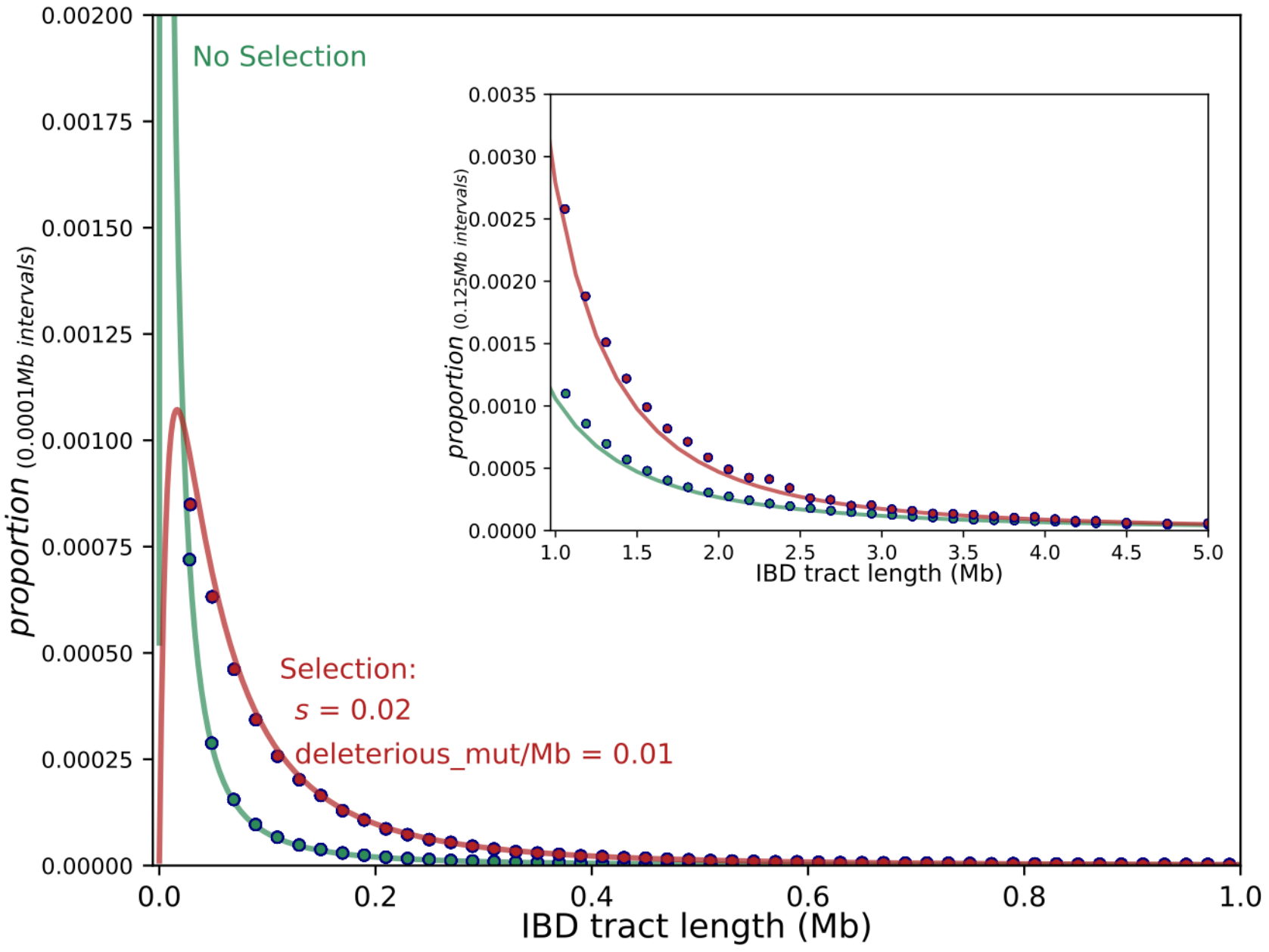

**Figure 3**. Effect of background selection on the distribution of IBD tract lengths in a random-mating population of 1,000 individuals. The distribution is expressed as the proportion of the genome covered by IBD tracts of a given length. The lines show theoretical predictions from Equation (10), computed using the program PDF_IBD_tracts_BS.py, and the dots show the corresponding simulation results. Green indicates the neutral expectation, whereas red indicates background selection. The recombination rate was $c = 0.01$ per Mb. The inset enlarges the 1–5 Mb range. As the input of new mutations was very high (approximately 30 times the recombination rate), mutation dominated the determination of tract boundaries.

## - The effects of selective sweeps on the distribution of IBD tract lengths

Selective sweeps operate locally and for a limited period, but their effects on linked neutral variation persist beyond the period of allele substitution. Gillespie (2000) showed that the pseudo-steady-state level of neutral variation under recurrent sweeps can be captured by a reduced $N_e$. Wang *et al.* (2016) subsequently demonstrated that Gillespie's expression for $N_e$ can be derived directly from the long-term contribution theory of Santiago and Caballero (1998).

In the general equation for $N_e$ under selection on an additive trait (Robertson 1961),

$$N_e = \frac{N}{1 + Q^2 V_w} \ ,$$

the “1” represents the variance of contributions due to the random sampling of gametes before the action of selection, and the term $[Q^2V_w]$ represents the variance of the expected long-term contribution of families attributable to selection.

With linkage, the equation does not hold for families nor diploid individuals. Instead, it refers to the $N_e$ of a neutral site in the genome, with $[Q^2V_w]$ being the variance of the expected long-term contributions of neutral copies from one generation projected into the future. Given a reference generation $t$ backwards in time at which the cumulative selective effect begins, and a final generation 0 at which the consequences of selection-induced drift are assessed, the variance of the long-term contributions of neutral copies from generation $t$ to generation *0* is given by:

$$[Q^2V_w]_{(-t)} = q_{-t} \cdot \left(\frac{y_0 q_0}{q_{-t}}\right)^2 + (1 - q_{-t})\left(\frac{1 - y_0 q_0}{1 - q_{-t}}\right)^2 - 1$$

where $q_{-t}$ and $q_0$ are the frequencies of the favourable allele at generations $t$ and 0 respectively, and $y_0$ is the frequency of the neutral copies at generation 0 that were initially associated with the favorable allele $t$ generations ago, expressed as a relative frequency within the set of chromosomes carrying the favorable allele (see Section 7 of the Appendix). The effective population size representing the genetic drift at generation $t$ projected on generation 0 is:

$$N_{e(-t)} = \frac{N}{1 + [Q^2V_w]_{(-t)}}$$

and the coalescence probability at generation $t$ of two copies sampled from generation 0 is

$$f(t) = \frac{1}{2N_{e(-t)}} \prod_{i=1}^{t-1} \left(1 - \frac{1}{2N_{e(-i)}}\right)$$

The effects of a selective sweep on the length distribution of IBD tracts are asymmetric with respect to the neutral focal site. Suppose the selected site is located to the left of the focal site. Recombination events on the right side of the focal site trim the right IBD boundary, but do not affect the association between neutral copies at the focal site and the selected allele. Consequently, the coalescence probabilities are independent of recombination events on the right, and can be combined with the PDF of IBD tract lengths to the right side of the neutral focal site ($P(x_R|t)$) in the standard way:

$$p(x_R; h) = h \cdot \sum_{t=1}^{\infty} [P(x_R|t) \cdot f(t)] \tag{11}$$

for a small interval of lengths $x_R \pm h/2$ Mb. Deriving the analogous expression for IBD tracts on the left side (i.e. the side containing the selected site) is substantially more complex. This is because recombination events in the interval between the selected and neutral sites alter both the degree of association with the favorable allele and the IBD tract boundary simultaneously. Therefore, the solution cannot be derived as a simple product of independent probabilities.

When the neutral site is located close to the selected site, as shown in the insert in Figure 4, tracts on the side containing the selected site tend to be longer than those on the opposite side. This asymmetry arises because tracts on the selected side that extend beyond the interval between the focal and selected sites share the selective conditions of the favorable allele. This shared advantage increases selection-induced drift, resulting in longer IBD tract lengths on that side. Consequently, this asymmetry between the two flanking distributions reveals which side of the neutral focal site the selected locus occupies.

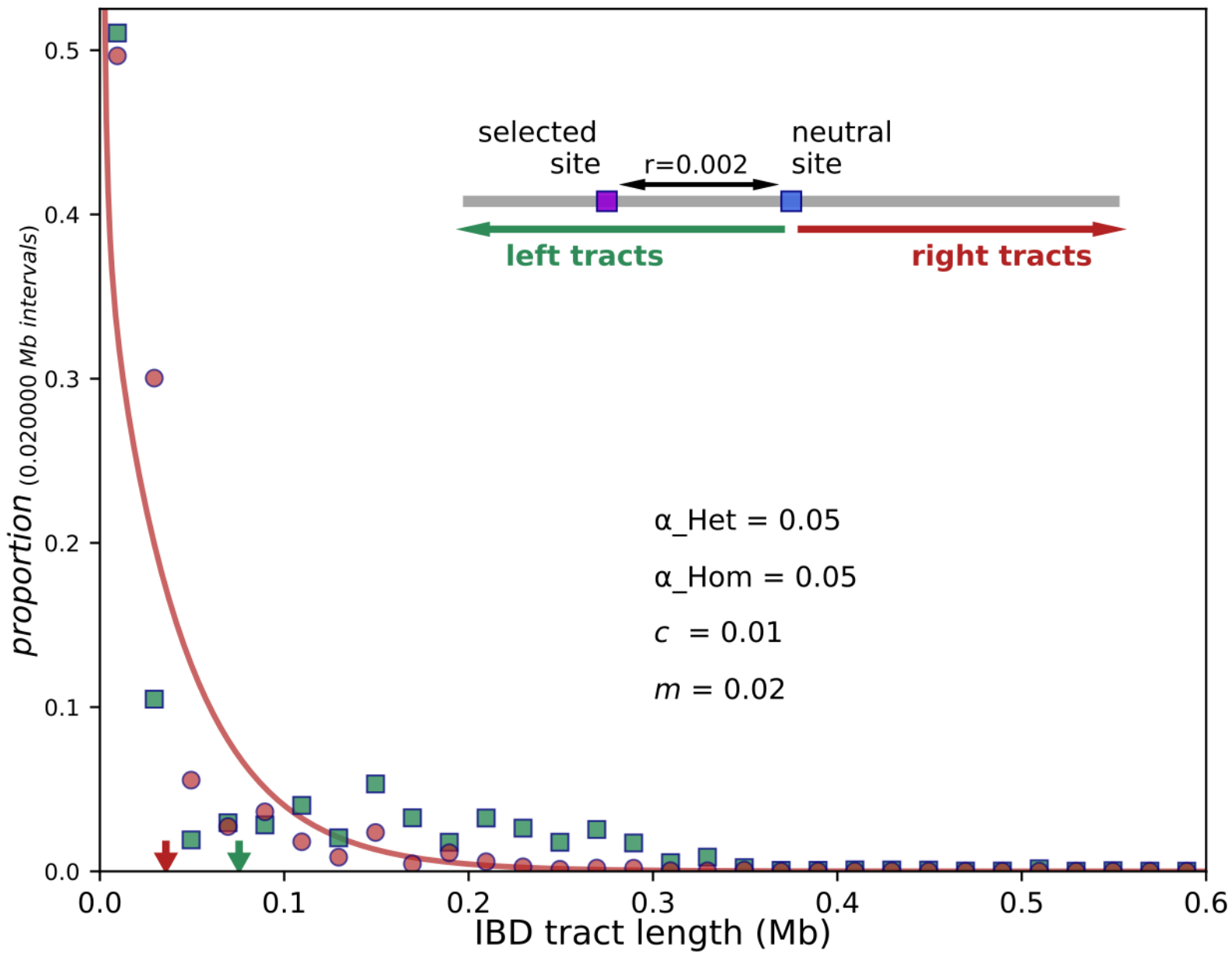

**Figure 4.** Effect of a selective sweep on IBD tract length distributions to the left and right sides of a neutral focal site. The focal site was located 0.2 Mb to the right of the favorable mutation. Recombination rate was 1 cM/Mb across the genome. Distributions were obtained from 200 individuals sampled when the favorable allele first reached a frequency of 90%. Red circles show the observed lengths of right-side tracts, extending away from the selected site, whereas green squares show those of left-side tracts, extending towards the selected site, as illustrated in the inset. The red line shows the theoretical prediction for right-side tracts obtained from Equation 11 using the program PDF_IBD_tracts_SS.py. Red and green arrows on the horizontal axis indicate the mean lengths of the right- and left-side distributions, respectively.

To track changes in the IBD tract length distribution during a simulated selective sweep, two local $N_e$ estimates were performed across the 3 Mb region surrounding the selected site (Figure 5): Equation A26, evaluated at the median tract length (i.e. at $p_{site}(x) = 0.5$), and Equation 5, based on the mean tract length. Both estimators identified the selected site as a local $N_e$ minimum but differed in sensitivity and persistency. The mean-based estimator produced an earlier, deeper signal, detecting the selective sweep when the favorable allele reached 20% frequency. In contrast, the median-based signal was initially weaker but rapidly sharpened and persisted longer. This is because the median, which is much smaller than the mean, is affected only by break events occurring within the short distance within the median value, while the mean integrates events over the full tract length distribution. Therefore recombination, mutation and gene conversion erodes the median-based signal very slowly, delaying its return to its initial level prior to the sweep.

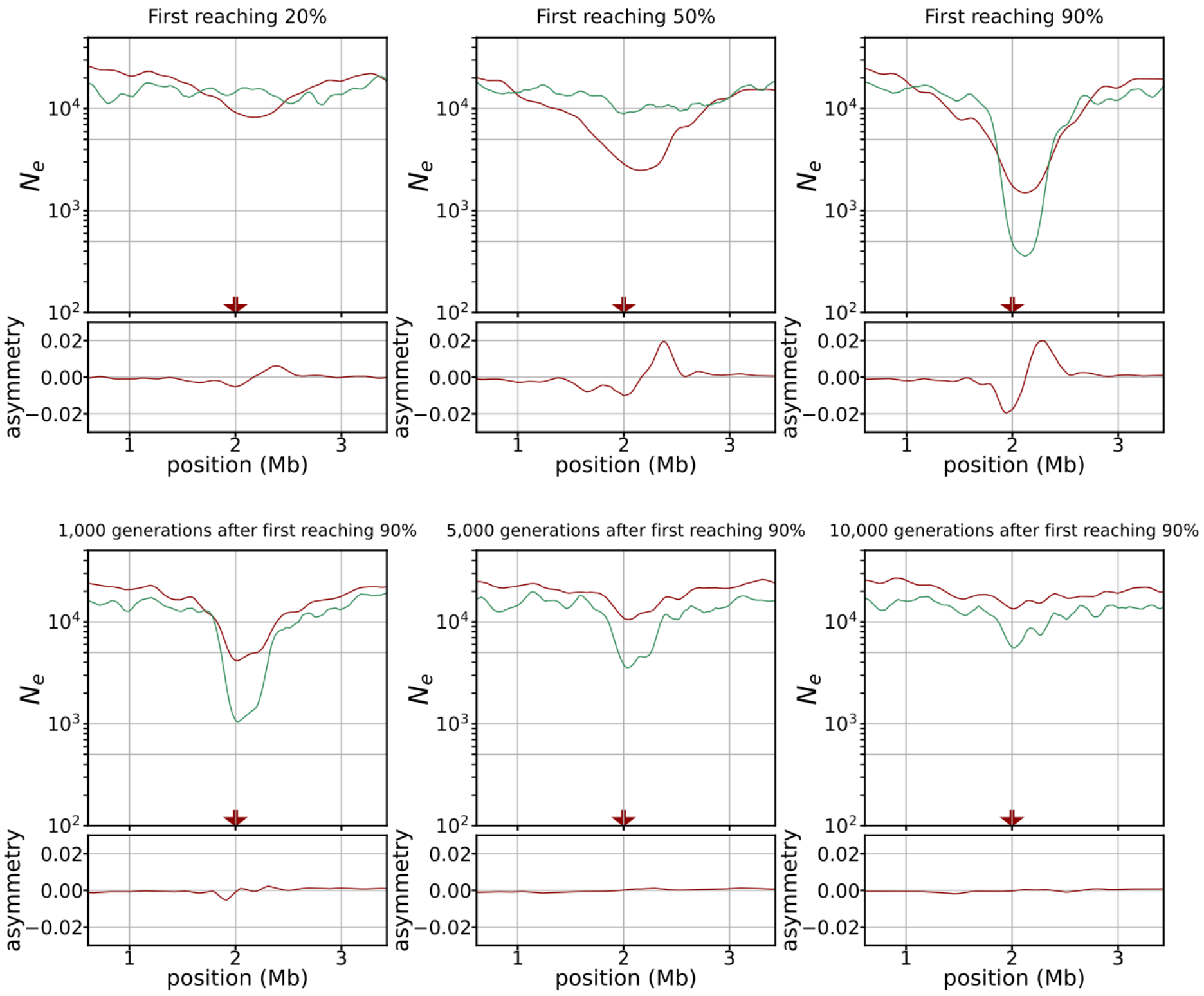

**Figure 5.** Temporal dynamics of a selective sweep inferred from IBD tract-length distributions. Each panel shows two local $N_e$ estimates (upper plot) and the left-right asymmetry in mean tract length (lower plot) across a simulated genomic region in a population of 15,000 individuals. Local $N_e$ was estimated using Equation A26 (green lines) and Equation 5 (red lines). Asymmetry was calculated as the mean left-side length minus the mean right-side length, in cM. Panels, ordered from left to right and top to bottom, correspond to the first time the favorable allele reached frequencies of 20%, 50%, and 90%, followed by 1,000, 5,000, and 10,000 generations after it first reached 90%. The favorable allele was dominant, with a selective advantage of $\alpha = 0.05$, and the recombination rate was $c = 0.01$ per Mb. Red arrows mark the position of the selected site.

The left-right asymmetry was measured as the difference between the means of the distributions on the left and right sides (Figure 5). A negative value (left-side length minus right-side length) indicates that the selected locus lies to the right of the neutral site, while a positive value indicates that it lies to the left. Although the left-right asymmetry was clearly evident during the selective sweep, its duration was short compared to the persistence of the local $N_e$ reduction. Therefore, combining both signals makes it possible to discriminate between ongoing or recent sweeps and ancient ones.

Two well-established examples of local adaptations of European populations were examined for local reductions in $N_e$ and left-right asymmetries around the corresponding genomic regions (Figure 6). Lactase persistence is associated with regulatory variants in the *MCM6* region (Enattah *et al.* 2002), whereas lighter skin pigmentation is strongly associated with the derived A111T variant of *SLC24A5* (Lamason *et al.* 2005; Ju and Mathieson 2021). The $N_e$ profiles were highly concordant between populations from the same continental group. In populations in which these adaptations are present, the candidate regions coincide with pronounced local $N_e$ minima at the locations of the reference genes (black arrows), particularly for the median-based $N_e$ estimates. These minima were absent or weaker at the corresponding positions in populations in which selection is presumed to have been absent or less intense. Left-right asymmetry provided a complementary signal, showing pronounced changes near the candidate genes in the populations that underwent adaptation.

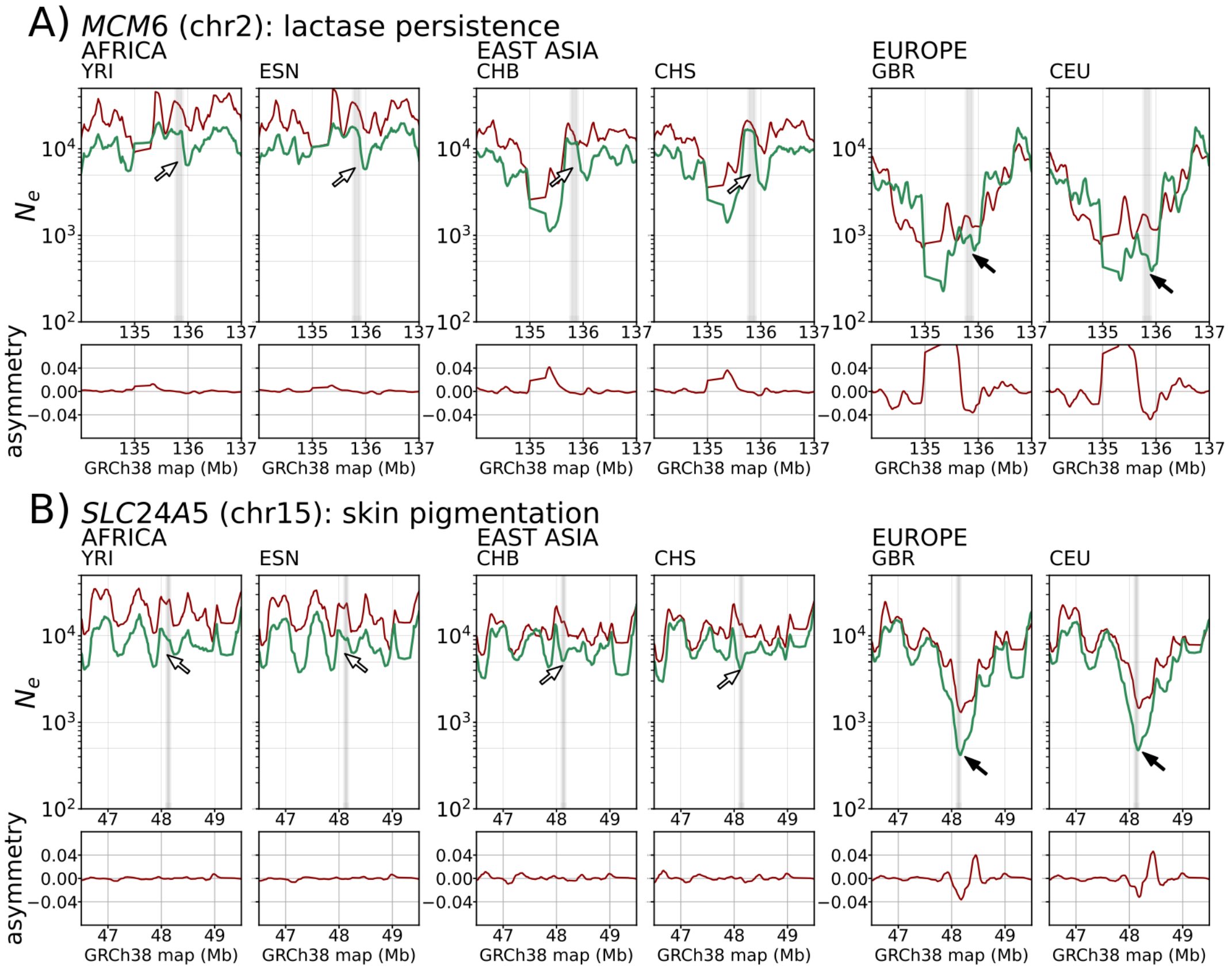

**Figure 6.** Local $N_e$ estimates and left-right asymmetries across two 3-Mb regions associated with human local adaptations. (A) The *MCM6* region associated with lactase persistence. (B) The *SLC24A5* region associated with skin pigmentation. Grey shading vertical bars show the positions of these genes. Six populations are grouped by continental origin: YRI and ESN (Africa), CHB and CHS (East Asia), and GBR and CEU (Europe). In each population, the upper plot shows local $N_e$ estimates along the two regions using Equation A26 (green lines) and Equation 5 (red lines). The lower plot shows the left-right asymmetry along the two regions. Black arrows indicate local $N_e$ minima using Equation A26 in populations with strong evidence of the selective sweep; white arrows mark the corresponding genomic positions in populations where the sweep is expected to be absent or weaker.

## - Extension to a time-varying effective population size

The time-varying-$N$ form of Equation 4 can be readily obtained by integrating Equation 2 over time, weighted by the coalescent probabilities $f(t)$ over all past generations:

$$p(x;h) = h \cdot \sum_{t=1}^{\infty} [P(x|t) \cdot f(t)] \tag{12}$$

$$f(t) = \frac{1}{2N_{(-t)}} \cdot \prod_{i=1}^{t-1} \left(1 - \frac{1}{2N_{(-i)}}\right)$$

where $N_{(-t)}$ is the effective population size at generation $t$ back in time and, by convention, the product term is taken to equal 1 when $i = 1$. This equation is implicit in Palamara *et al.* (2012) and Browning and Browning (2015) without accounting for mutation and gene conversion.

Inversion of Equation 12 to recover the past $N_{(-t)}$ series is difficult because a demographic trajectory has one unknown $N$ for each generation and there would be far more unknowns than independent pieces of information, i.e. bins of tract lengths. Many different histories would produce practically the same spectrum of tract lengths, and small errors in the observed coverages would be strongly amplified in the estimated population sizes. Grouping generations into epochs of constant size reduces the number of unknowns to what the data can resolve. A closed form solution when $N$ is assumed to be piecewise constant over consecutive generations is derived in Section 5 of the Appendix.

The past is divided into $K$ epochs of constant population size, and the coverage of each length bin is expressed as a sum of contributions from these epochs (Equations A14 and A16). A bin of length $x$ receives a contribution from epoch $k$ only if tracts of that length can remain unbroken until the start of the epoch. This occurs with a probability of $e^{-\sigma}$, where $\sigma = 2Bx\tau_{k-1}$ is the expected number of break events accumulated before the epoch begins, and $\tau_{k-1}$ is the last generation of the previous epoch. Therefore, long tracts inform only on the recent past and shorter tracts reach progressively further back. This allows the epochs to be resolved sequentially, from the most recent to the oldest. For each new epoch, the contributions of the epochs already resolved are subtracted from the observed coverage of the corresponding bin. The remainder, rescaled by the probability of reaching that epoch uncoalesced and unbroken, is denoted $R_k$. Treating the epoch as extending indefinitely into the past, the expectation of this remainder depends on the population size of the epoch, $N_k$, only through $w = 1/(1 + 4BN_kx_k)$. This is the probability that a tract of length $x_k$ is not broken before coalescence, given that the two lineages enter the epoch uncoalesced. Equating this expectation to the observed remainder $R_k$ gives a cubic equation for $w$ (Equation A17), which is solved explicitly, and $N_k$ is obtained directly from $w$ (Equation A18). The resulting $N_k$ is not the population size at a particular generation, but rather a weighted harmonic average of $N_{(-t)}$ over a period centered around generations of the order of $1/(Bx_k)$. Figure 7 compares the true demographic trajectory of a simulated population with the inferred piecewise trajectory obtained by applying Equation A18 using the Ne_trajectory.py program.

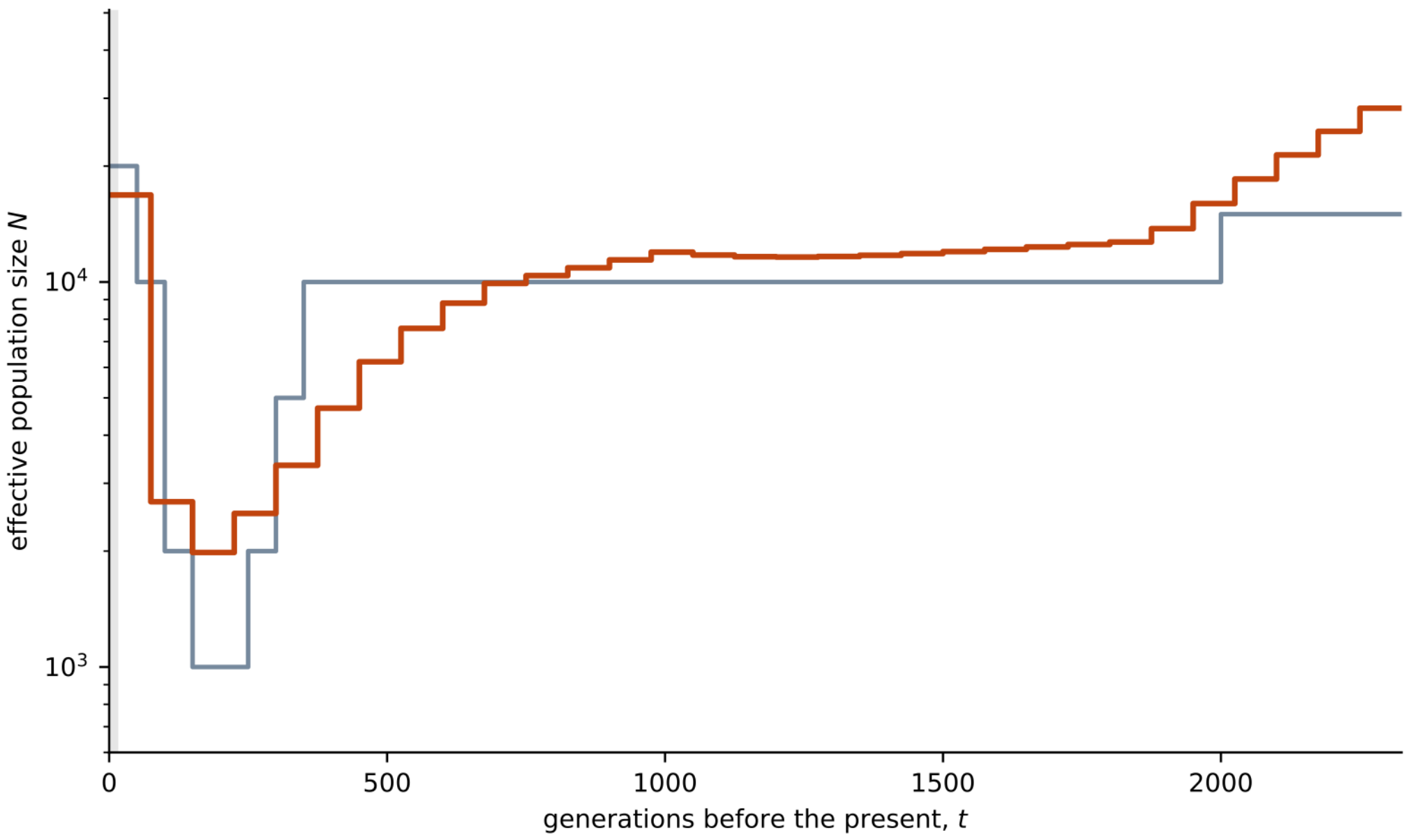

**Figure 7.** Reconstruction of a piecewise-constant trajectory of effective population size from the observed ROH length distribution. The blue line shows the true simulated trajectory, and the orange line shows the trajectory inferred sequentially from Equation A18 using Ne_trajectory.py with epochs of 75 generations. Each inferred value represents the harmonic mean of $N_{(t)}$ over the corresponding epoch. Simulated genomes comprised ten 150-Mb chromosomes, with a constant recombination rate of $c = 0.01$ per Mb (i.e. 1 cM/Mb) and a mutation rate of 2.5 $10^{-8}$ per site per generation.

# Discussion

## - From IBD tracts to observable ROH

The two tract-length distributions considered here differ only in their sampling scheme. A tract spanning a specified focal site consists of two independent flanking lengths, each exponentially distributed. Their sum therefore follows an Erlang distribution with a mean twice that of tracts randomly sampled from all genomic IBD tracts, that is, twice the mean distance between consecutive tracts across the genome. The factor of two arises not from the coalescent process itself, but from the length-biased sampling method: a focal site is more likely to fall within a long tract than a short one, in proportion to the genomic length covered by each tract. Consequently, the focal-site distribution is the length-biased version of the inter-break interval distribution (Chapman and Thompson 2003; Tiret and Hospital 2017).

What this implies for inference is less often made explicit. The two sampling schemes yield different estimators of $N$ and different expected coalescence times. For an IBD tract sampled by spanning a focal site, without conditioning on its length, the expected coalescence time remains $2N$ generations, consistent with Kingman's (1982) pairwise coalescent. For a tract sampled at random from all IBD tracts in the genome, the expectation is $4N$, because removing the focal-site condition gives greater representation to the more numerous short tracts. For very short tracts, the expectation approaches $6N$, reflecting the enrichment for ancient coalescence events and the long period over which breakpoints must accumulate to reduce the segment to such short lengths. For long tracts, the expected coalescence time becomes independent of $N$.

The link from theoretical IBD tracts to empirically observed ROHs was stablished in two steps. First, mutation and gene conversion were incorporated as additional break processes whose rates add to the recombination rate. Unlike recombination, mutations and gene conversions do not cause conflation, which reduces the relative impact on short-tract inference. As mutation and conversion rates are usually given in terms of physical distance, the framework adopted a physical map rather than a genetic one. This eliminates the propagation of errors from the genetic map to the mutation and conversion rates. Second, the effect of finite marker density was modelled by assuming that heterozygous sites occur independently along the chromosome. Under this assumption, the displacement from the true recombination breakpoint to the nearest detectable heterozygous site follows a hypoexponential distribution. Although the model ignores correlations in heterozygosity along chromosomes, its predictions closely match simulations across a wide range of population sizes, provided that marker density is sufficiently high.

The CSH statistic (Sved 1971; Hayes *et al.* 2003) emerges as a special case: it is the probability that an IBD tract spans both the focal site and a second site at a given distance. The original formulation does not treat mutation and gene conversion as sources of identity breakage. Equation 6 incorporates these events as breakpoints of identity and Equation A26 adds the consideration of finite marker density.

## - Resolution and limits of demographic inference from tract lengths

The demographic history can be recovered from the tract-length spectrum because each length bin is a different window to the past: long tracts survive unbroken only if coalescence was recent, whereas short tracts reach progressively further back. This ordering allows a history divided into epochs to be resolved one epoch at a time, in closed form and without iterative fitting. The coalescence density at a particular length $x$ Mb follows an Erlang distribution with shape parameter 3, and therefore the upper limit of the 90% central interval is 6.30/0.82=7.7 times its lower limit, which is located at $0.82/(2xB + 1/2N)$ generations. Thus, each bin averages the coalescence density over a wide window of time and the number of epochs that can be estimated is bounded by the information in the data. Each $N$ estimate is a harmonic average of the population size over its window. Uncertainty also accumulates towards the past, since each epoch is resolved after subtracting the contributions of the more recent ones.

A tract of length $x$ Mb provides information about coalescence times of the order of $1/(2Bx)$ generations. This establishes the minimum temporal limit of inference, which is determined by the length of the longest tracts in the data. The upper limit is determined by the accuracy with which short tracts can be measured. These limitations position tract-based inference between two well-established families of methods. Linkage disequilibrium between linked sites with a recombination frequency of $r$ reflects drift over approximately $1/2r$ generations. As pairs of loosely linked sites are abundant across the genome, methods based on linkage disequilibrium (e.g. Santiago *et al.* 2025) can resolve the last hundred generations. Methods based on the density of heterozygous sites, such as PSMC and MSMC (Li and Durbin 2011; Schiffels and Durbin 2014), rely on mutations that accumulate in proportion to coalescence time. Consequently, they provide information about the distant past but observe few recent mutations and therefore little signal in recent coalescences. Tract lengths combine recombination and mutation as break events, so they bridge the two timescales.

One caveat applies to all coalescent-based methods, including this one. What is recovered is the inverse of the instantaneous coalescence rate of two lineages, which equals the effective size only in panmictic models. Any process that alters the rate at which two lineages coalesce, such as variable population structure and non-random mating, leaves the same signature as a genuine change in size (Rodriguez *et al.* 2018).

## - Background selection and the IBD tract length distribution

Since Robertson (1961) it has been understood that selection on heritable traits accelerates drift by inducing correlations in allele-frequency change across consecutive generations that a single round of meiosis does not erase. Santiago and Caballero (1998, 2016) showed that the consequence cannot be represented by a single reduced $N_e$: drift is more intense for older neutral mutations than for recent ones, so the relevant quantity is a generation-dependent $N_{e(t)}$, indexed by the age of the mutation. That formulation has since been used to predict the spectrum of neutral variation (Buffalo and Kern 2024; Becher and Charlesworth 2025), an effect that the standard background-selection model of Charlesworth *et al.* (1993) does not represent. What has not been done before is to carry it into the length distribution of IBD tracts. Substituting $N_{e(t)}$ into Equation 10 shows that background selection shifts the distribution towards longer tracts and that no constant $N_e$ reproduces its shape, because the effective size governing recent coalescence is larger than the one governing ancient coalescence. This is the direct IBD-tract analogue of the effect of selection on nucleotide diversity (Santiago and Caballero 2016), and it places the quantitative-genetic theory of selection and the IBD coalescent within a single framework.

## - Selective sweeps: localization, persistence and asymmetry

Selective sweeps transiently distort the distribution of IBD tract lengths at sites linked to the selected locus. These distortions were predicted by adapting the long-term contribution theory (Robertson 1961; Santiago and Caballero 1995) to the sweep scenario and integrating it with the IBD tract length theory developed here. This generalizes the approach restricted to tracts centered on the selected site itself (Temple *et al.* 2024), to neutral focal sites throughout the surrounding genomic region. The framework also introduces the left-right asymmetry in tract length as a directional signal: its sign indicates on which side of a focal site the selected locus lies.

The two local $N_e$ estimators developed here provide complementary information on the sweep. The mean-based estimator (Equation 5) responds earlier because it integrates information across the full tract-length distribution and is therefore sensitive to perturbations over multiple genomic distances. Consequently, it can detect the sweep while the favorable allele is still at low frequency. By contrast, the median-based estimator (Equation A26) initially produces a weaker signal, but this signal strengthens rapidly, localizes the selected site more precisely, and persists for many generations after fixation. Left-right asymmetry provides additional temporal information: it is strongest during and shortly after the sweep but decays rapidly thereafter, making it particularly informative about recent sweeps.

## - Concluding remarks

This work connects IBD tract-length distributions and empirically observed ROH length distributions within a single coalescent framework. Mutation, gene conversion, and marker density enter directly into the model rather than being treated as corrections applied afterwards. The framework also accommodates time-varying population size, allowing closed-form reconstruction of piecewise-constant demographic histories. In addition, it represents selection-induced reductions in effective population size as a function of lineage age rather than as a single constant $N_e$. This contribution is primarily theoretical, providing a foundation for inference methods that remain to be developed: likelihood-based estimation of population size using the full tract-length distribution, models incorporating population structure, and statistically calibrated tests for localizing and dating selective sweeps.

## Data and Software Availability Statement

The author affirms that all data necessary for confirming the conclusions of the article are present within the article and its Appendix. Quantitative predictions are validated against computer simulations using the parameters described in the Materials and Methods section and in the main text. The following Python programs used to compute the theoretical predictions and to generate simulated tract length distributions are freely available at https://github.com/esrud/ROHvsIBD/:
PDF_IBD_tracts_BS.py predicts IBD tract-length distributions under background selection.
PDF_IBD_tracts_SS.py predicts IBD tract-length distributions throughout a selective sweep.
Ne_trajectory.py estimates the demographic trajectories from a table of ROH frequencies.

## Acknowledgements

The author thanks Ino Curik and Luboš Vostrý for their helpful comments on an earlier version of the manuscript, Carlos Köpke for providing scripts to analyze data and John Sved for a clarifying exchange on one of the key equations.

**The author declares no competing interests.**

# Appendix

Index:

## 1- The general model

The Wright-Fisher model is assumed, consisting of a population of $2N$ haploid individuals with sexual reproduction. The genome is treated as a continuous sequence into which an infinitesimally small focal site is located. This focal site is virtual and has no associated genotype.

At generation $t = 0$, two descendants from the same individual receive replicate copies of the focal site, which are transmitted along two independent genealogical branches across subsequent generations (Figure A1). Recombination along each branch progressively trims the original chromosome on both sides of the focal site, so that only the genomic region immediately flanking the focal site is retained in the present generation.

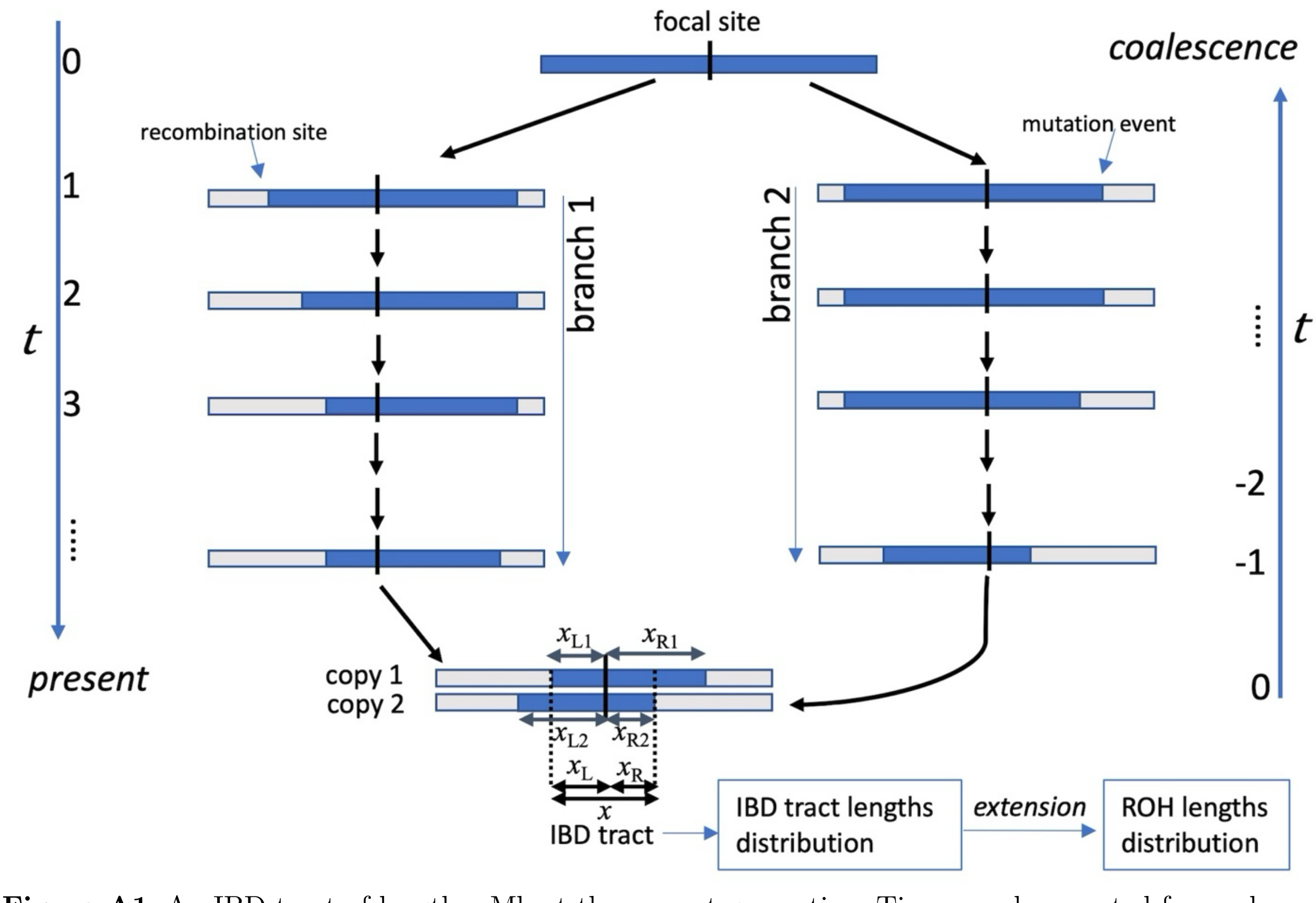

**Figure A1**. An IBD tract of length $x$ Mb at the present generation. Time may be counted forward in positive numbers or backward in negative numbers. Recombination progressively trims both copies (copy 1 and copy 2) of the original chromosome on either side of the focal site, defining left (L) and right (R) boundaries of the IBD tract. Because recombination breakpoints are unobservable, the empirically detectable entity is an ROH, whose boundaries are defined by heterozygous marker sites located beyond the true IBD limits.

The region of overlap between the two retained copies defines an IBD tract associated with the focal site. This tract is bounded on each side by the most recent recombination breakpoints on the respective branches (Figure A1). Since recombination breakpoints cannot be observed directly from sequence data, the IBD tract cannot be precisely measured. Instead, it is observed as a run of homozygosity (ROH) whose boundaries are displaced outwards from the true recombination sites to the nearest heterozygous marker sites. The model also incorporates *de novo* mutations and gene conversion events, which interrupt the identity continuity similarly to recombination.

## 2- The two distributions of IBD tract lengths conditional on coalescence time

In addition to recombination, occurring at a rate $c$ per Mb, mutation and gene conversion break the continuity of identity at a combined rate $m$ per Mb. Assuming there is no interference between break events, the total number of breaks per Mb follows a Poisson distribution with rate $B = c + m$. The distances between consecutive breaks therefore follow an exponential distribution. This is the same distribution as the distance from the fixed focal point to the nearest breakpoint on either side. In the first generation, the PDF of the distance $x_{R1}$ Mb from the focal site to the nearest breakpoint on the right side of branch 1 is $B \cdot e^{-Bx_{R1}}$.

As breaks accumulate independently across generations, the total rate over $t$ generations is $tB$, and the PDF of lengths on the right-hand side of the focal site is exponential as well:

$$P(x_{\mathrm{R1}}|t) = tB \cdot e^{-tB \cdot x_{R1}}$$

The right IBD tract boundary, at $x_{\mathrm{R}}$ Mb from the focal site, is set by the shorter of the two branch distances $x_{\mathrm{R1}}$ and $x_{\mathrm{R2}}$ (Figure A1). Because the minimum of two independent exponential variables with rate $tB$ is itself exponential with rate $2tB$, the PDF of $x_{\mathrm{R}}$ is equivalent to the single-branch distribution evaluated at twice the number of generations:

$$P(x_{\mathrm{R}}|t) = P(x_{\mathrm{R1}}|2t) = 2tB \cdot e^{-2tB \cdot x_{\mathrm{R}}} \tag{A1}$$

The left and right flanking tract lengths are independent and share the same exponential distribution. Therefore, the sum of these lengths, $x = x_{\mathrm{L}} + x_{\mathrm{R}}$, follows an Erlang distribution with shape parameter 2 and rate parameter $2tB$:

$$P(x|t) = \mathrm{Erlang}_{2,2tB} = 4t^2B^2 \cdot x \cdot e^{-2tB \cdot x} \tag{A2}$$

All raw moments of this distribution are available in closed form. In particular, the expected length of an IBD tract that includes the focal site is:

$$E(x)_P = \frac{1}{tB} \text{ Mb} \tag{A3}$$

This is half of the value obtained by Fisher (1949) and Hanson (1959) for the mean length of a heterozygous segment around a selected gene in a backcrossing program. The difference arises because breaks accumulate along a single chromosome in the backcross setting, whereas in the coalescent framework they accumulate independently along two branches, effectively doubling the rate of boundary erosion. Although Equation A2 assumes an infinitely long chromosome, this is not a practical limitation: breaks accumulate rapidly on both sides and in both branches. Additionally, autosomes are typically at least 0.5 morgans long. Therefore, deviations from the infinite-chromosome approximation are negligible even for the most recent coalescence events (see Table 1 in Hanson 1959).

Equation A2 describes the conditional distribution of IBD tract lengths given that the tract containing the focal site, as opposed to the overall distribution of IBD tract lengths across the genome. The two distributions differ because the set of tracts containing the focal site is a length-biased subsample (see Figure A2): the probability that a randomly chosen tract spans the focal site is proportional to its length.

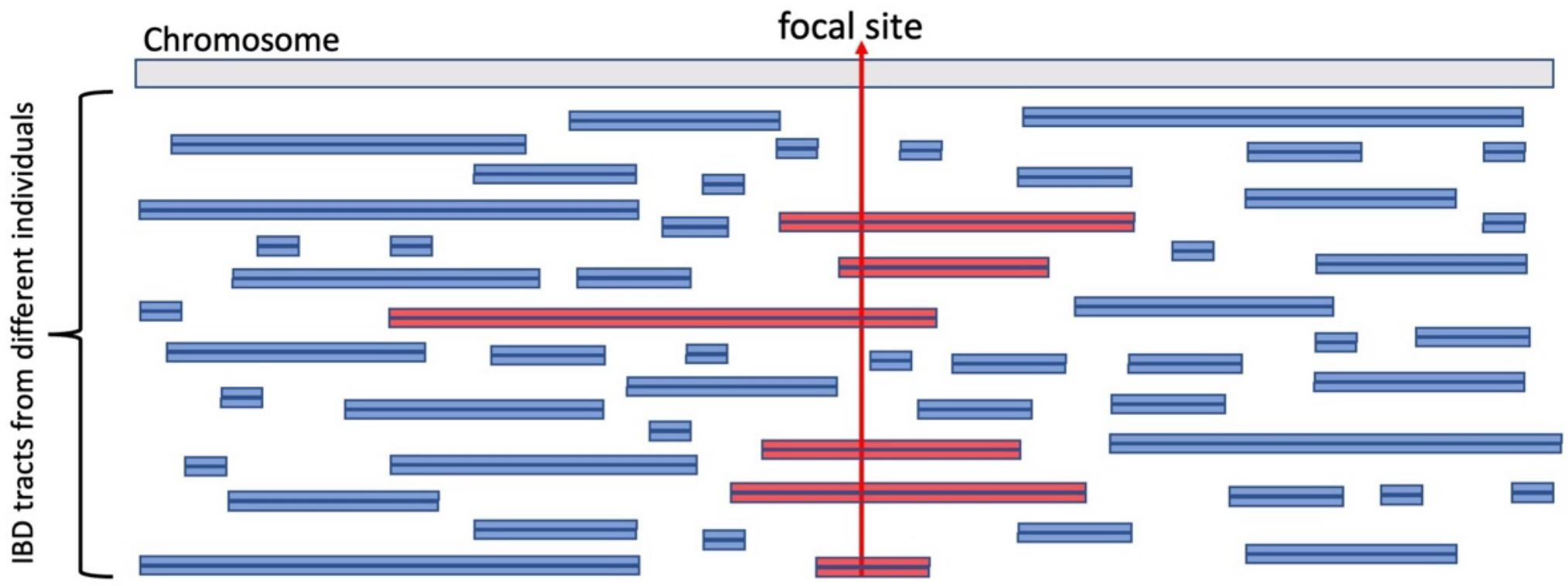

**Figure A2**. Simulated positions of IBD tracts on a chromosome in a sample of individuals. Each tract is represented by a horizontal double bar representing the two homologous segments. A focal site is located at the center of the chromosome. Tracts spanning this focal site are shown in red. Because the probability that a randomly chosen tract contains the focal site is proportional to its length, the red subset constitutes a length-biased sample of the full tract distribution.

The genome-wide distribution of IBD tract lengths is therefore proportional to $P(x|t)/x$:

$$\frac{P(x|t)}{x} = 4t^2B^2 \cdot e^{-2tB \cdot x}$$

This expression is not a proper PDF since it does not integrate to unity. After normalization, the overall PDF tract length distribution becomes:

$$Q(x|t) = \frac{P(x|t)/x}{\int_0^\infty P(x|t)/x \cdot dx} = 2tB \cdot e^{-2tB \cdot x} \tag{A4}$$

which is an exponential distribution, identical to Equation A1, with mean:

$$E(x)_Q = \frac{1}{2tB} \quad \text{Mb} \tag{A5}$$

This mean is half that of Equation A3, reflecting the length bias introduced when conditioning on the focal site.

## 3- The distribution of coalescence times of IBD tracts conditional on tract length

Unlike the distributions derived in the previous section, the distribution of coalescence times conditional on observed IBD tract length depends on the effective population size $N$. In finite populations, the density of $P(x|t)$ (Equation A2) decreases backwards in time due to the declining probability of coalescence at earlier generations. Throughout the following derivations, exact coalescence probabilities of the form $f(t) = \frac{1}{2N}\left(1 - \frac{1}{2N}\right)^{t-1}$ are used wherever a closed-form solution is attainable; the approximation $\frac{1}{2N}e^{-t/2N}$ is used where it is not. Multiplying Equation A2 by the coalescence probability and normalizing to a proper PDF yields the distribution of coalescence times for all sites contained within IBD tracts of length $x$ Mb:

$$R(t|x) = \frac{P(x|t) \cdot f(t)}{\int_0^\infty P(x|t) \cdot f(t) \cdot dt} = \frac{1}{2}t^2 \left(2xB + \frac{1}{2N}\right)^3 \cdot e^{-t\left(2xB+\frac{1}{2N}\right)}$$

(A6)

Note that this distribution applies to tracts of a given length regardless of whether they contain a specific focal site. Equation A6 has the form of an Erlang distribution with shape parameter 3 and rate parameter $\varphi = 2xB + 1/(2N)$ and, consequently, all the raw moments are available in closed form. In particular, the mean coalescence time of a tract of length $x$ Mb is:

$$E(t) = \frac{3}{2xB + \frac{1}{2N}} \text{ generations}$$

(A7)

For any Erlang distribution of shape 3, the coefficient of variation is $1/\sqrt{3} = 0.577$ and the 90% central interval ranges from $0.82/\varphi$ to $6.30/\varphi$ generations regardless of the rate. For the longest tracts, the expectation becomes essentially independent of population size and approaches $3/(2xB)$ generations. By contrast, for very short tracts the resulting mean coalescence time approaches $6N$ generations, well above the standard expectation of $2N$ generations under Kingman's (1982) pairwise coalescence model. The expected coalescence time of a tract randomly sampled from those containing the focal site is $2N$ generations, as shown by averaging the expectations of Equation A7:

$$\int_0^\infty P(x|t) \cdot E(t) \cdot dx = 2N \text{ generations}$$

However, the expectation for a tract randomly sampled anywhere in the genome, unconditionally of containing any focal site, is double the latter expectation:

$$\int_0^\infty Q(x|t) \cdot E(t) \cdot dx = 4N \text{ generations}$$

## 4- Equations for $N$ in terms of the observed IBD tract lengths

Consider first the steady-state distribution $P(x)$ of IBD tract lengths containing a focal site in a genome at mutation-recombination-drift equilibrium with no selection, treating all genome sites as equivalent. Integrating the product of the PDF of tract lengths conditional on the coalescence time (Equation A2) and the coalescence probability over all the coalescence times gives:

$$P(x) = \int_0^\infty P(x|t) \cdot f(t) \cdot dt = \frac{4xB^2}{N\left(2xB + \frac{1}{2N}\right)^3}$$

This equation also represents the proportion $p_{IBD}(x)$ of the genome covered by IBD tracts of length *x* Mb: dividing the equation by *x* first converts the focal-site tract length density to the overall tract length density, and multiplying by *x* again yields the genome coverage by IBD tracts of length *x*:

$$p_{IBD}(x) = \left(P(x) \cdot \frac{1}{x}\right) \cdot x = P(x)$$

For a window of width *h* Mb centered at *x*, the fraction of the genome covered by IBD tracts within that size range is:

$$p(x; h) = \frac{4xB^2 \cdot h}{N \cdot \left(2xB + \frac{1}{2N}\right)^3} \tag{A8}$$

When $x$, expressed in Mb, is much greater than $(4BN)^{-1}$, this simplifies to:

$$p(x; h) \approx \frac{h}{2N \cdot x^2 B} \quad \Rightarrow \quad N \approx \frac{h}{2p(x; h) \cdot x^2 B}$$

This expression allows $N$ to be estimated directly from the observed proportion of the genome covered by IBD tracts of a given length. The corresponding coalescence times for these tracts are given by Equation A7. Because shorter tracts coalesce deeper in the past, applying the estimator across a range of tract lengths reveals temporal variation in $N$.

While Equation A8 provides a genome-wide average $N$, site-specific estimates are valuable for detecting the effects of local selection on genetic drift. Two complementary local estimators are derived below.

The first is based on the mean length $\bar{x}_{site}$ of IBD tracts containing a focal site. Summing the expected tract length conditional on coalescence time (Equation A3), weighted by the coalescence probability, yields:

$$\bar{x}_{site} = \sum_{t=1}^{\infty}\left[\frac{1}{tB}\cdot f(t)\right] = \frac{\ln(2N)}{2NB} \tag{A9}$$

The second estimator is based on the proportion $p_{site}(x)$ of IBD tracts that contain both the focal site and a reference site at a distance of $x$ Mb, equivalently the proportion of tracts with lengths $\geq x$. For a single chromosome, the probability of no break between the focal and reference sites is the Poisson 0 class probability $e^{-xB}$. For two chromosomes that coalesced $t$ generations ago, this probability is $e^{-2txB}$. Summing over all coalescence times gives:

$$p_{site}(x) = \sum_{t=1}^{\infty}[\, e^{-2txB}\cdot f(t)] = \frac{1}{2N(e^{2xB}-1)+1} \approx \frac{1}{4NxB+1} \tag{A10}$$

Unlike Equation A9, which integrates information over the full chromosome length, Equation A10 depends only on events occurring within a distance of $x$ Mb from the focal site, making it more sensitive to detect local drift variations. In practice, a target proportion is chosen (e.g. $p_{site}(x) = 0.5$ for the median), the corresponding distance $x$ is identified from the sample data, and $N$ is solved from Equation A10.

The expected coalescence time for IBD tracts spanning both the focal and the reference sites is obtained by averaging the coalescence times, each weighted by a summand of Equation A10:

$$t_{site|x} = \frac{\sum_{t=1}^{\infty}[\mathrm{t}\cdot e^{-2txB}\cdot f(t)]}{\sum_{t=1}^{\infty}[e^{-2txB}\cdot f(t)]} = \frac{1}{(1-e^{-2xB})+\frac{1}{2N}} \approx \frac{1}{2xB+\frac{1}{2N}} \tag{A11}$$

## 5- Extension to time-varying population size with piecewise-constant epochs

The expressions derived above assume a constant population size. This section relaxes that assumption. It first writes the length distributions for an arbitrary demographic history and then shows that, if the history is approximated by a sequence of epochs of constant size, the size of each epoch can be obtained in closed form from the observed spectrum of tract lengths.

Let $N_{(-t)}$ be the effective population size $t$ generations before present. In each generation t, two lineages coalesce with probability $1/2N_{(-t)}$. The probability that they have not coalesced by generation $t$ is therefore $S(t) = e^{-\sum_{i=1}^{t-1}(1/2N_{(i)})}$, and the probability of coalescence at generation $t$ is

$$f(t) = \frac{1}{2N_{(-t)}} \cdot e^{-\sum_{i=1}^{t-1}\left(\frac{1}{2N_{(-i)}}\right)}$$

(A12)

Every tract-length quantity derived so far is an average over the coalescence time, of a term of the form $e^{-2tBx}$ multiplied by a power of $t$. This is because the probability that a tract of length $x$ is not broken in $t$ generations on two branches is $e^{-2tBx}$. Averages of this kind are the Laplace transform of $f(t)$ and its derivatives, evaluated at $s = 2xB$:

$$f_L(s) = \int_0^{\infty} e^{-st} \cdot f(t) \cdot dt \approx \sum_{t=0}^{\infty} e^{-st} \cdot f(t)$$

(A13)

Note that this equation is equivalent to Equation A10, that is the probability that an IBD tract spans two sites at a distance $x$ apart, now for an arbitrary history.

The coverage of a length bin can be written in the same terms. Substituting Equation A2 into Equation 12 and taking outside the sum every factor that does not depend on $t$ gives $p(x;h) = 4hB^2x \cdot \sum t^2 e^{-st} f(t)$. Differentiating $f_L(s) = \sum e^{-st} \cdot f(t)$ twice with respect to $s$ results in a sum that is the second derivative of the transform, $\sum t^2 e^{-st} f(t) = f_L''(s)$, so that

$$p(x;h) = 4hB^2x \cdot f_L''(s)$$

(A14)

Each observed length bin provides one value of $f_L''$ at one value of $s$ set by its length. Short tracts correspond to small $s$ and look far back in time; long tracts correspond to large $s$ and look at the recent past. In the same way, averaging Equations A1, A4 and A10 over $f(t)$ gives $p_{site}(x) = f_L(s)$ for any history, and Equations A5 and A9 generalize to $f_L'(s)$ and $f_L''(s)$ respectively.

Because $S(t)$ declines only by coalescence, $f(t) = -S'(t)$. Therefore, knowing the density $f(t)$ would determine the history:

$$2N_{(-t)} = \frac{S(t)}{f(t)} = -\frac{S(t)}{S'(t)}$$

(A15)

This identity is exact, but it cannot be applied directly. The data do not provide $f(t)$ at individual generations, only values of its transform, each of which averages $f(t)$ over a broad range of times. Recovering $f(t)$ from its transform is numerically unstable: small errors in the observed coverages produce large errors in the reconstructed history. The approach taken here is to restrict the history to a small number of epochs of constant size, which makes the problem stable and with solution in closed form, as shown below.

Divide the past into $K$ epochs. Epoch $k$ spans the generations between $\tau_{k-1}$ and $\tau_k$, with $\tau_0 = 0$ and $\tau_K = \infty$, and has constant size $N_k$. Within this epoch, lineages coalesce at rate $\lambda_k = 1/(2N_k)$. Let $\Delta_k = \tau_k - \tau_{k-1}$ be the duration of the epoch and let $S_{k-1} = \exp\left(-\sum_{j<k} \lambda_j \Delta_j\right)$ be the probability that two lineages reach the beginning of the epoch $k$ without having coalesced. Integrating $f(t)$ epoch by epoch gives

$$f_L(s) = \sum_{k=1}^{K} S_{k-1} \frac{\lambda_k}{s + \lambda_k} \cdot e^{-s\tau_{k-1}} \cdot \left(1 - e^{-(s+\lambda_k)\Delta_k}\right) \tag{A16}$$

a finite sum of $K$ exponential terms, one per epoch. With a single epoch, this is the constant-$N$ case of Sections 1-4. The discrete sum of Equation 12 is recovered by replacing $e^{-t/2N}$ with $\left(1 - \frac{1}{2N}\right)^t$.

Because the coverage of a length bin is proportional to $f_L''(s)$ (Equation A14), the coverage expected under a piecewise-constant history is obtained by differentiating Equation A16 twice with respect to $s$:

$$f_L''(s) = \sum_k \lambda_k S_{k-1} e^{\lambda_k \tau_{k-1}} [\Phi(\tau_{k-1}; s + \lambda_k) - \Phi(\tau_k; s + \lambda_k)]\,,$$

$$\Phi(t; a) = e^{-at} \left(\frac{t^2}{a} + \frac{2t}{a^2} + \frac{2}{a^3}\right)$$

with $\Phi(\tau_K; a) = 0$.

The factor $e^{-s\tau_{k-1}}$ in Equation A16 determines how much epoch $k$ contributes to a bin of length $x$. Let $\sigma = 2Bx\tau_{k-1}$ be the expected number of breaks accumulated in the two lineages over tracts of length $x$ during the $\tau_{k-1}$ generations preceding epoch $k$. The factor $e^{-s\tau_{k-1}} = e^{-\sigma}$ in Equation A16 is the probability that a tract of length $x$ survives unbroken until epoch $k$ begins. It is close to 1 when $\sigma$ is small and falls below 5% once $\sigma$ exceeds about 3. Long tracts are broken before reaching ancient epochs, so they carry information only on the recent past. Bins can therefore be ordered from long to short, and each new bin used to determine one more epoch, further back in time, once the more recent ones are known.

When epoch $k$ is being resolved, it is treated as the last one, extending indefinitely into the past ($\Delta_k \to \infty$). Its term in Equation A16 then reduces to $T_k(s) = S_{k-1} e^{-\sigma} w$, where

$$w = \frac{\lambda_k}{s + \lambda_k} = \frac{1}{1 + 4BN_k x}$$

is the probability that a tract of length $x$ is not broken before the two lineages coalesce, given that they entered epoch $k$ uncoalesced (Equation A10 with constant $N$).

The observed coverage of any bin contains the contributions of all epochs. Let $G_{<k}$ be the part of $f_L''(s)$ due to the epochs already determined, that is, the sum above restricted to $j < k$. Subtracting it from the observed value gives the contribution of epoch $k$. Dividing that contribution by the known factors $S_{k-1}e^{-\sigma}$ and by $2/s^2$ to make it dimensionless, gives

$$R_k = \frac{s^2}{2} \cdot \frac{p(x;h)/(4hB^2x) - G_{<k}}{S_{k-1}e^{-\sigma}}.$$

$R_k$ is the part of the observed coverage of bin $x$ that remains to be explained by epoch k. It remains to equate $R_k$ with the second derivative of $T_k(s)$:

$$T_k''(s) = S_{k-1}e^{-\sigma}\frac{w}{s^2}[\sigma^2 + 2\,\sigma\,(1-w) + 2(1-w)^2]$$

Setting the dimensionless $\frac{s^2}{2}T_k''(s)/(S_{k-1}e^{-\sigma})$ equal to $R_k$:

$$w^3 - (\sigma+2)w^2 + \frac{(\sigma+1)^2+1}{2} \cdot w - R_k = 0 \tag{A17}$$

Then $w$ is the root of this cubic equation. Once $w$ is known, the definition of $w$ above is solved to obtain the effective size of the epoch:

$$N_k = \frac{(1-w)}{4Bxw} \tag{A18}$$

Equation A17 is solved by Cardano's formula. Writing it as $w^3 + b_2w^2 + b_1w + b_0 = 0$, with $b_2 = -(\sigma+2)$, $b_1 = [(\sigma+1)^2+1]/2$ and $b_0 = -R_k$, and setting $p = b_1 - b_2^2/3$ and $q = 2b_2^3/27 - b_1b_2/3 + b_0$,

$$w_n = -\frac{b_2}{3} + 2\sqrt{-\frac{p}{3}} \cdot \cos\left[\frac{1}{3}\arccos\left(\frac{3q}{2p}\sqrt{-\frac{3}{p}}\right) - \frac{2\pi n}{3}\right], \qquad n = 0, 1, 2$$

where $n$ labels the three roots of the cubic equation. When $4p^3 + 27q^2 > 0$ there is a single real root, and the cosine is replaced by the corresponding hyperbolic function.

Because $w$ is a probability, the valid solution is the root that lies in the interval (0, 1) and where two of them do, the valid solution is the larger $N_k$. For $\sigma > 2$ only one root lies in (0, 1), and the ambiguity is confined to the first epoch. For the first epoch $\sigma = 0$, $S_0 = 1$ and $G_{<1} = 0$, so that $R_1 = p(x;h)\,x/(2h)$ and Equation A17 reduces to $w^3 - 2w^2 + w - R_1 = 0$, a function of the observed bin.

The quantity returned for epoch $k$ is not $N_{(-t)}$ at any particular generation. It is the constant size that, applied from $\tau_{k-1}$ onwards, reproduces the coverage observed at the length $x_k$. In practice, $N_k$ is the harmonic average of $N_{(-t)}$ over $t > \tau_{k-1}$, with weights proportional to $t^2e^{-2Bx_kt}$. This weight increases with $t^2$ but is reduced by the exponential, so it concentrates on times of order $1/(Bx_k)$: each epoch is therefore reported as a smooth average over a limited window, and abrupt changes within that window are flattened.

The program Ne_trajectory.py, available at https://github.com/esrud/ROHvsIBD/, estimates $N_k$ (Equation A18) from observed frequencies across specified ROH length bins.

## 6- The distribution of IBD tract lengths under background selection

Consider a background selection model at mutation-selection-drift balance and effects distributed uniformly over the genome. Also consider a large number of replicated populations, each of $2N$ haploid individuals. Neutral copies of two alternative alleles are randomly assigned to the focal site across all replicates at generation $t = 0$, so that every population begins with the same reference allele frequency $p$.

As the replicated populations reproduce independently, neutral allele frequencies diverge progressively due to genetic drift. From the perspective of an observer at $t = 0$, the rate of between-population divergence in gene frequency ($\Delta V$) is constant over time when selection is absent (Figure A3). The rate of increase in the within-population inbreeding coefficient ($\Delta F$) mirrors this divergence, and both equal $1/(2N_e)$ per generation

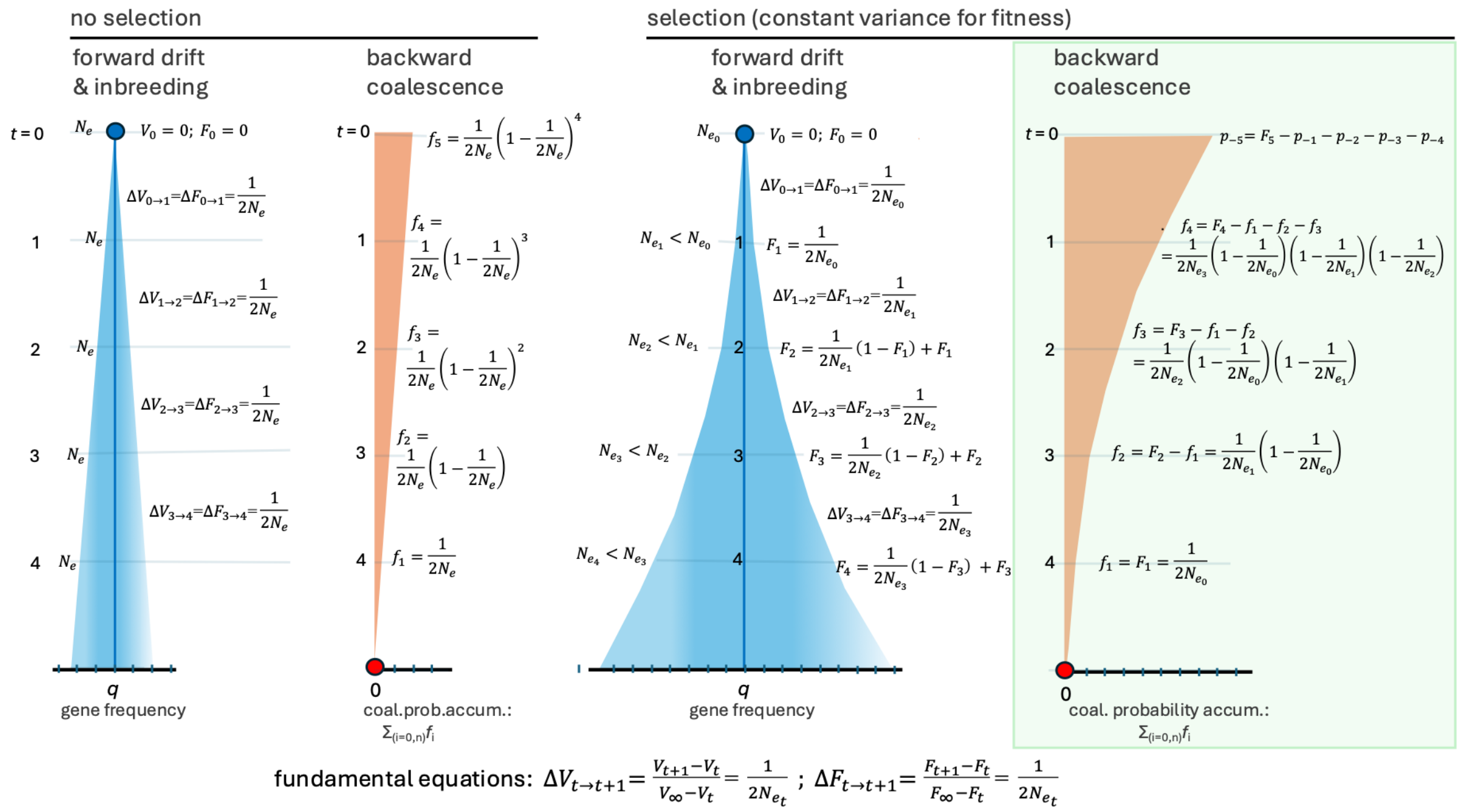

**Figure A3.** Between-population variance in gene frequency ($V$), within-population inbreeding coefficient ($F$), and pairwise coalescence probability ($f_t \coloneqq f(t)$) for neutral alleles, under scenarios with and without selection. At $t = 0$, neutral copies are randomly distributed at frequency $q$ across individuals of replicated populations. Populations are assumed to have reached mutation-selection-drift balance prior to $t = 0$. Starting from $t = 0$ (blue dots), the effective population size at each generation ($N_{e(t)}$), is estimated recursively from either $V$ or $F$ using the fundamental equations shown.

In the absence of selection, $N_e$ remains constant. Under selection, $N_e$ decreases progressively until an asymptotic value is reached, reflecting the cumulative effect of heritable variance for fitness in reproductive success. In the pairwise coalescence perspective (right panels), the reference point shifts to the generation at which two neutral copies are assessed for IBD (red dots). The probabilities of coalescence at preceding generations depend on the same series of consecutive $N_{e(t)}$ values but traversed in reverse chronological order.

Under selection, positive correlations between successive gene frequency changes cause both the rate of divergence and the rate of inbreeding to increase across generations until a limiting value is reached (Figure A3). This limiting state corresponds to the asymptotic $N_e$ first identified by Robertson (1961) in the context of artificial selection on quantitative traits with unlinked polygenes, and later extended by Santiago and Caballero (1995, 1998) to non-random mating and natural selection at linked loci. Under a multiplicative background selection model (Charlesworth *et al.* 1993), the increasing drift and inbreeding effects on neutral mutations are fully captured by a decreasing series of consecutive $N_e$ values indexed from the generation in which a neutral mutation arises ($t = 0$) (Santiago and Caballero 2016):

$$N_{e(t)} = N \cdot e^{-\frac{2}{cL}\int_0^{cL/2} V_w \cdot Q_{r(t)}^2 dr} \tag{A19}$$

where $Q_{r(t)} = \sum_{i=0}^{t}(1-r)^i \cdot \left(1 - \frac{V_M}{V_w}\right)^i$ represents the cumulative effect of selection over $t$ generations, $L$ is the chromosome length in Mb, so that, at a constant rate of $c$ recombination events per Mb, $cL$ is its length in morgans, $V_w$ is the standing genetic variance for fitness and $V_M$ is the input of new genetic variance for fitness per generation. Note that $N_{e(t)}$ is relative to a virtual time-point defined by the origin of a new mutation and cannot be assigned to a specific calendar generation of the population.

This framework allows background selection to be incorporated into Kingman's (1982) coalescent model. Two key considerations underpin the approach. First, the probability of coalescence one generation ago ($f_1$ in Figure A3) depends only on the population size and variance in family size at that generation; it has not accumulated a selective effect over time, so:

$$f_1 = \frac{1}{2N_{e(0)}}$$

Second, the inbreeding coefficient at any generation equals the sum of all pairwise coalescence probabilities in preceding generations. In the notation of Figure A3:

$$F_t = \sum_{i=1}^{t} f_i$$

Under these two conditions, the probability of pairwise coalescence $t$ generation ago is:

$$f_t = \frac{1}{2N_{e(-t)}} \prod_{i=1}^{t-1} \left(1 - \frac{1}{2N_{e(-i)}}\right)$$

where the $N_e$ values are taken from Equation A19.

The selective effect on coalescence probabilities accumulates backwards in time from the present to the generation of coalescence. Coalescence events in the most recent generations are only weakly affected by selection; as one looks further back in time, the associated $N_e$ values decrease, and the corresponding coalescence probabilities rise above the neutral expectation. These coalescence probabilities can be combined with the PDFs of IBD tract lengths conditional on the coalescence time (Equation A2 for $P(x|t)$), provided that the recombination events that shape the LD between the selected and neutral sites are uncorrelated with those trimming IBD tract boundaries. Under this assumption, the proportion of the genome covered by IBD tracts of lengths within the interval $x \pm h/2$ Mb is (following the derivation of Equation A8):

$$p_{IBD}(x;h) = h \cdot \sum_{t=1}^{\infty} [P(x|t) \cdot f(t)] \tag{A20}$$

If the density of selective effects is assumed uniform across the genome, this equation applies to focal sites located anywhere in the genome, with the exception of sites close to telomeric ends, where the effective density of linked selected sites is reduced.

The program PDF_IBD_tracts_BS.py, available at https://github.com/esrud/ROHvsIBD/, computes $p_{IBD}(x;h)$ (Equation A20) from user-specified parameters.

## 7- The distribution of IBD tract lengths under positive selection

Unlike the background selection model, in which fitness variance is distributed uniformly across the time and the genome, a selective sweep is a discrete event localized in a particular genome site and compressed in time, even though its effects on linked neutral variation persist well beyond the period of allele substitution. Gillespie (2000) showed, using a pseudo-hitchhiking model in which favorable mutations recur at a site, that the steady-state level of neutral variation can be captured by a reduced $N_e$. Wang *et al.* (2016) subsequently demonstrated that Gillespie's expression for $N_e$ follows directly from the long-term contribution theory of Santiago and Caballero (1998). The derivation below extends that theory to predict the distribution of IBD tract lengths during a partial or complete selective sweep at a single site. The general equation for $N_e$ under selection on an additive trait is (Robertson 1961):

$$N_e = \frac{N}{1 + Q^2 V_w} \ ,$$

where $V_w$ is the variance of expected contributions of the subjects (with mean 1, i.e. the squared coefficient of variation of expected contributions to the next generation), that is the variance for fitness, and $Q$ is the cumulative effect of selection. Under random mating without selection, $V_w = 0$ as all the subjects are expected to contribute equally to the next generation; for selection on a non-heritable trait, $Q = 1$, so that only the immediate variance of contributions matters. Together, $Q^2V_w$ captures the variance in expected long-term contributions attributable to selection on an inherited trait. The subjects of the equation can be families (Robertson 1961) or diploid individuals (Santiago and Caballero 1995) when loci are unlinked; with linkage, the relevant units are haplotypes or chromosomes, and the equation gives the $N_e$ associated with a specific neutral site (Santiago and Caballero 1998, 2016). For multiplicative fitness effects the equation takes the more convenient form $N_e = N \cdot e^{-Q^2V_w}$.

Let $q$ be the frequency of a favorable allele in a population of $2N$ haploids, with selective advantage $\alpha$ assigned from the moment the allele first appeared by mutation as a single copy. The trajectory of a favourable allele destined for fixation can be decomposed into two epochs (Santiago and Caballero 2005). From frequency $1/(2N)$ to $1/(2N\alpha)$, the copy number follows a branching process with an expected increase of one copy per generation. From $1/(2N\alpha)$ until near fixation, the trajectory closely follows the deterministic expectation. The final approach to fixation is drift-dominated and its precise timing is irrelevant for linked neutral variation. For the purpose of predicting $N_e$, the expected trajectory is therefore well approximated by the first two epochs. The same decomposition applies to diploids with allele action from additivity to partial dominance, with the heterozygous effect $\alpha$ being the relevant parameter for delimiting the two epochs.

Let $r$ be the recombination frequency between the selected site and a neutral focal site and let $y_n$ be the frequency at generation $n$ of a particular class of neutral copies within the set of chromosomes carrying the favorable allele. Given that the favorable allele follows the expected trajectory, the expected frequency $y_{n+1}$ in the next generation of the neutral copies associated with the favorable allele in the previous generation, and the expected squared frequency $y^2_{n+1}$, evolve according to the recursions:

$$y_{n+1} = y_n \cdot (1 - (1 - q_n)r)$$

$$y^2_{n+1} = y^2_n \cdot (1 - 2(1 - q_n)r) + \frac{y_n - y^2_n}{2N \cdot q_n}$$

where the second term in the latter equation accounts for sampling within the set of chromosomes carrying the favorable mutation.

Two temporal points of reference are defined: the present generation, or generation 0, at which the consequences of the selective sweep are assessed, and a generation t backwards in time from which the cumulative effect of selection begins. Let $q_{-t}$ and $y_{-t} = 1$ be the initial frequencies of the favourable allele and of the class of neutral copies associated with it (i.e. the class is defined by the association with the favourable allele $t$ generations ago), and $q_0$ and $y_0$ their expected values at present (after the sweep or partial sweep). Both $y_{-t}$ and $y_0$ are relative frequencies within the set of chromosomes carrying the favourable allele. Each of the $2Nq_t$ neutral copies initially linked to the favourable allele is expected to reach $y_0q_0/q_{-t}$ copies after the sweep, while each of the remaining $2N(1-q_{-t})$ neutral copies are expected to reach $(1-y_0q_0)/(1-q_{-t})$ copies. The variance of the expected long-term contributions of neutral copies from generation $t$ to the present is therefore:

$$[Q^2V_w]_{-t} = q_{-t}\cdot\left(\frac{y_0q_0}{q_{-t}}\right)^2 + (1-q_{-t})\left(\frac{1-y_0q_0}{1-q_{-t}}\right)^2 - 1 \ ,$$

and the effective population size representing the genetic drift at generation $t$ projected on generation 0 is:

$$N_{e(-t)} = \frac{N}{1+[Q^2V_w]_{(-t)}}$$

The coalescence probability $t$ generations ago of two copies sampled at generation 0 is

$$f_t = \frac{1}{2N_{e(-t)}}\prod_{i=1}^{t-1}\left(1-\frac{1}{2N_{e(-i)}}\right)$$

A selective sweep is a strong, local perturbation whose effect on the IBD tract length distribution is asymmetric: it depends on which side from the neutral focal site contains the selected locus. Recombination events in between the neutral and selected sites (say on the left side from the neutral site) break down the LD between the two sites, thereby modulating the selection-induced drift at the neutral locus and, consequently, the probabilities of coalescence. Because recombination events on the left and right sides of the focal site are independent, the coalescence probabilities can be combined with the PDF ($P(x_R|t)$) of IBD tract lengths to the right side of the neutral focal site in the standard way:

$$p_{IBD}(x_R;h) = h\cdot\sum_{t=1}^{\infty}[P(x_R|t)\cdot f(t)] \tag{A21}$$

as before, for a small interval of lengths $\pm h/2$ Mb around $x_R$.

The derivation of the analogous PDF for IBD tracts on the side that includes the selected site is substantially more complex. Recombination events within the interval between the selected and neutral sites simultaneously affect both the association with the favorable allele and the tract boundaries, so the solution is not a simple product of independent probabilities.

The program PDF_IBD_tracts_SS.py, available at https://github.com/esrud/ROHvsIBD/, computes $p_{IBD}(x_R;h)$ (Equation A21) for user-specified parameters.

## 8- The effect of finite marker density on the PDF of ROH lengths

In the above derivations, the boundaries of the IBD segments are defined by recombination, mutation or gene conversion sites. In practice, however, samples consist of a finite number of marker sites, meaning that the ROH tracts identified empirically do not coincide with the IBD segments. Strictly speaking, the equations above cannot be applied directly to real observations. One way to address this issue is to incorporate the uncertainty in the displacement of boundaries directly into the model.

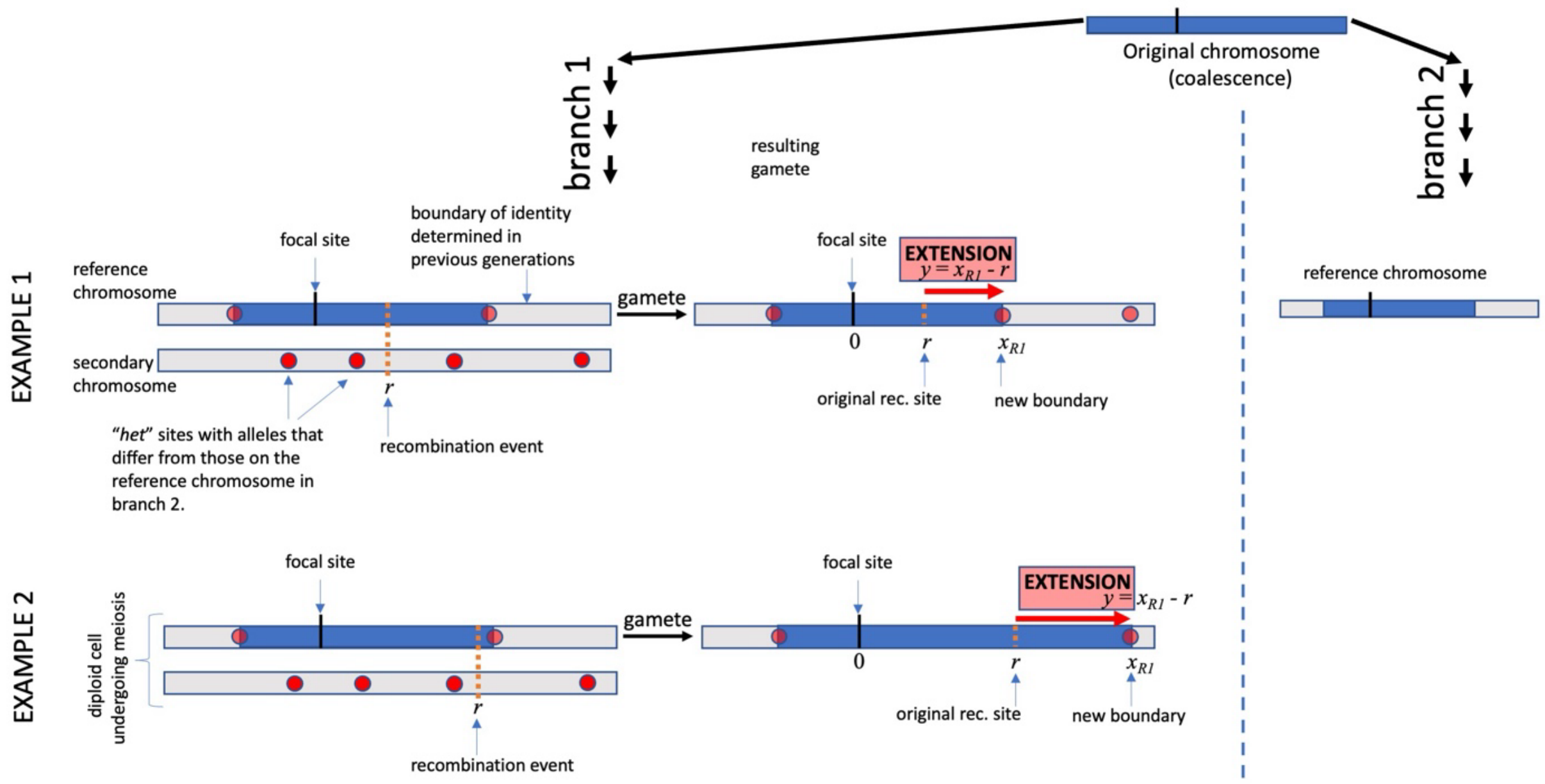

**Figure A4**. Two examples of identity extensions beyond the recombination breakpoints. Identities between the two reference chromosomes of the two branches are shown in blue color, and *het* sites (i.e. sites with different alleles in both reference chromosomes) are marked by red circles. In Example 1, the new identity limit with respect to the chromosome in branch 2 extends beyond the recombination site and moves to the first *het* site, resulting in a slight reduction in the identity segment. In Example 2, the new boundary extends beyond the previous one, resulting in a longer identity segment. The magnitude $y$ of the extensions is equal to the difference between the true recombination site $r$ and the location $x_{R1}$ of the *het* site that defines the new identity boundary. The subscript R1 refers to the right-hand side of the focal site and branch 1 of the genealogy (see Figure A1).

The following derivation adopts the notation introduced in Figures A1 and A4. Consider a genomic region in which *het* sites (sites heterozygous between the two random chromosomes) occur at an average rate of $H/D$, where $H$ is the average heterozygosity of segregating sites and $D$ the average physical distance in Mb between adjacent segregating markers. Assuming *het* sites are randomly distributed along the chromosome, their counts follow a Poisson distribution, and the PDF of the extension length $y$ (in Mb) is exponential with rate $\lambda_1 = H/D$:

$$\frac{H}{D}e^{-y\frac{H}{D}}$$

Similarly, the distance *r* in Mb between consecutive true recombination breakpoints follows an exponential distribution with rate $\lambda_2 = c$ recombination events per Mb:

$$c \cdot e^{-rc}$$

The observable right-hand boundary of the conserved segment from the focal site is determined by two additive components: the distance *r* to the nearest recombination breakpoint and the extension *y* beyond it to the nearest *het* site. Since both components are exponentially distributed, their sum $x = r + y$ follows a hypoexponential distribution with rates $\lambda_1$ and $\lambda_2$:

$$\text{probability function} = \frac{\lambda_1 \cdot \lambda_2}{\lambda_1 - \lambda_2}\left(e^{-x\lambda_2} - e^{-x\lambda_1}\right) = \frac{c}{1-\gamma} \cdot \left(e^{-xc} - e^{-x\frac{H}{d}}\right)$$

where $\gamma = cD/H$.

The probability that the boundary distance exceeds a given length $x_{R1}$ Mb:

$$\int_{x_{R1}}^{\infty} \frac{c}{1-\gamma} \cdot \left(e^{-xc} - e^{-x\frac{H}{d}}\right) dx = \frac{1}{1-\gamma} \cdot \left(e^{-cx_{R1}} - \gamma e^{-x_{R1}\frac{H}{d}}\right)$$

Now consider the additional contribution of mutation and gene conversion breaks, which occur at a combined rate *m* per Mb per generation and are Poisson distributed independently of recombination. In one generation, the probability that branch 1 shows no break of any kind (recombination with extension, mutation or gene conversion), within a distance $x_{R1}$ from the focal site, is the product of the respective survival probabilities:

$$P(x_{R1}) = \frac{1}{1-\gamma}\left(e^{-cx_{R1}} - \gamma \cdot e^{-x_{R1}\frac{H}{d}}\right) \cdot e^{-x_{R1}m} \tag{A22}$$

Across *t* generations of break accumulation, this probability becomes $P^t_{(x_{R1})}$. The corresponding PDF of the distance to the first observable break is obtained by differentiating the complement. Dropping the negligible cross-terms, yields the simpler approximation:

$$P(x_{\mathrm{R1}}|t) \approx e^{-tx_{R1}B}\left[\frac{1}{1-\gamma}\right]^t \cdot t \cdot \left[B - c \cdot e^{-x_{R1}\left(\frac{H-cD}{D}\right)}\right] \tag{A23}$$

Extending to both genealogical branches, the length $x_{\mathrm{R}}$ of the ROH to the right of the focal site equals the minimum of $x_{\mathrm{R1}}$ and $x_{\mathrm{R2}}$, the observable boundary distances on branches 1 and 2 respectively. That is, the PDF of $x_{\mathrm{R}}$ is equivalent to the one-branch distribution evaluated at twice the number of generations:

$$P(x_{\mathrm{R}}|t) = P(x_{\mathrm{R1}}|2t) = e^{-2tx_R B}\left[\frac{1}{1-\gamma}\right]^{2t} \cdot 2t \cdot \left[B - c \cdot e^{-x_R\left(\frac{H-cD}{D}\right)}\right]$$

When marker density is infinite ($D \to 0$, so $H/(cD) \to \infty$), the extension $\gamma \to 0$ and this expression reduces to Equation A1, in which case IBD and ROH boundaries coincide.

The total ROH tract length $x = x_R + x_L$ is the sum of independent and identically distributed left and right flanking lengths. Substituting $x_L$ by $(x - x_R)$ and the PDF for $x_L$ takes the following form:

$$P(x_{\mathrm{L}}|t) = e^{-2t(x-x_R)B}\left[\frac{1}{1-\gamma}\right]^{2t} \cdot 2t \cdot \left[B - c \cdot e^{-(x-x_R)\left(\frac{H-cD}{D}\right)}\right]$$

The solution for the PDF of the total length $x$ can be calculated as the integral of the product of $P(x_{\mathrm{R}}|t)$ and $P(x_{\mathrm{L}}|t)$:

$$P(x|t) = \int_{x_{\mathrm{R}}=0}^{x} P(x_{\mathrm{R}}|t) \cdot P(x_{\mathrm{L}}|t) \cdot dx_{\mathrm{R}}$$
$$= 4t^2 \cdot e^{-2txB}\left[\frac{1}{1-\gamma}\right]^{4t} \cdot \left[xB^2(1-\gamma) - 2B\gamma + [c^2x(1-\gamma) + 2B\gamma]e^{-x\left(\frac{H-cD}{D}\right)}\right]$$

This reduces to Equation A2 when $D \to 0$. For $x \gg D/H$, that is for tracts much longer than the mean physical distance between het sites, the expression simplifies to a form analogous to the pure IBD case with a modified effective rate, yielding the closed-form approximation:

$$P(x|t) \approx 4t^2B^2 \cdot x \cdot e^{-2txB} \cdot \left[\frac{1}{1-\gamma}\right]^{4t} = 4t^2B^2 \cdot x \cdot e^{-2txB} \cdot e^{-4t \cdot \ln(1-\gamma)} \tag{A24}$$

Combining this probability conditional on time of coalescence with the probability of pairwise coalescence at the focal site, a closed-form expression for the distribution of ROH lengths is obtained:

$$P(x) = \int_0^{\infty} P(x|t) \cdot \frac{e^{-\frac{t}{2N}}}{2N} \cdot dt = \frac{4xB^2}{N\left(2xB + \frac{1}{2N} - 4\gamma\right)^3}$$

From this equation, the fraction of the genome covered by ROH tracts within a small window of width $h$ centered at length $x$ is:

$$p(x; h) = h \cdot P(x) \tag{A25}$$

The ROH length distribution to the right side of the focal site conditional on time of coalescence is:

$$P(x_R|t) \approx 2tB \cdot e^{-2txB}\left[\frac{1}{1-\gamma}\right]^{2t}$$

Finally, a closed-form expression for the proportion $p_{site}(x)$ of ROH tracts that include both the focal site and a reference site at distance $x$ Mb (the ROH analogue of Equation A10) can be derived from Equation A22. The probability of no break of any kind between the focal and reference sites on two chromosomes that coalesced $t$ generations ago is $P_{(x_{R1})}^{2t}$. Integrating over coalescence times gives:

$$p_{site}(x) = \int_{t=0}^{\infty} P_{(x_{R1})}^{2t} \cdot \frac{e^{-\frac{t}{2N}}}{2N} \cdot dt = \frac{1}{1 - 2N\ \ln(X)} \tag{A26}$$

where $X = \left(\frac{e^{-xB} - \gamma e^{-x\frac{H}{D}}}{1-\gamma}\right)^2$

This reduces to Equation A10 when the density of markers is infinite.